\documentclass[fleqn,usenatbib]{mnras}

\usepackage{tensor} 

\usepackage[dvipsnames,hyperref]{xcolor}
\usepackage{multirow}
\hypersetup{
    pdftitle={MACS1149-JD1 Marconcini JWST},
    }

\usepackage{mathptmx}
\usepackage{amsmath}
\usepackage{xspace}
\newcommand{\msun}{\ensuremath{\mathrm{M_\odot}}\xspace}
\newcommand{\zsun}{\ensuremath{\mathrm{Z_\odot}}\xspace}
\newcommand{\mstar}{\ensuremath{\mathrm{M_\star}}\xspace}
\newcommand{\kms}{\ensuremath{\mathrm{km\,s^{-1}}}\xspace}
\newcommand{\peryr}{\ensuremath{\mathrm{yr^{-1}}}\xspace}

\def\arcsec{$^{\prime\prime}$}
\def\arcmin{$^{\prime}$}

\newcommand{\JWST}{\textit{JWST}\xspace}

\newcommand{\Spitzer}{\textit{Spitzer}\xspace}

\let\oldAA\AA
\renewcommand{\AA}{\text{\oldAA}\xspace}
\let\oldmicron\micron
\renewcommand{\micron}{\text{\oldmicron}\xspace}
\let\oldarcsec\arcsec
\renewcommand{\arcsec}{\text{\oldarcsec}\xspace}

\usepackage{xargs}
\usepackage{xspace}
\usepackage{textgreek}

\newcommand{\Lyalpha}{\text{Ly\textalpha}\xspace}
\newcommand{\Halpha}{\text{H\textalpha}\xspace}
\newcommand{\Hbeta}{\text{H\textbeta}\xspace}
\newcommand{\Hgamma}{\text{H\textgamma}\xspace}
\newcommand{\Hdelta}{\text{H\textdelta}\xspace}

\newcommand{\Hepsilon}{\text{H\textepsilon}\xspace}
\newcommand{\Hzeta}{\text{H\textzeta}\xspace}

\newcommandx{\permittedEL}[6][1=O,2=III,3=,4=,5=,6=]{\text{{#1}\,{\sc {#2}}{#3}{#4}{#5}{#6}}\xspace}
\newcommandx{\semiforbiddenEL}[6][1=O,2=III,3=,4=,5=,6=]{\text{{#1}\,{\sc{#2}}]{#3}{#4}{#5}{#6}}\xspace}
\newcommandx{\forbiddenEL}[6][1=O,2=III,3=,4=,5=,6=]{\text{[{#1}\,{\sc{#2}}]{#3}{#4}{#5}{#6}}\xspace}

\newcommandx{\NVL}[1][1=1243]{\permittedEL[N][v][\textlambda][#1]}
\newcommandx{\NVall}{\permittedEL[N][v][\textlambda][\textlambda][1239,][1243]}

\newcommandx{\CIIL}[1][1=232x]{\semiforbiddenEL[C][ii][\textlambda][#1]}
\newcommandx{\CIIall}{\semiforbiddenEL[C][ii][\textlambda][\textlambda][2324--][2329]}

\newcommandx{\NIVL}[1][1=1486]{\semiforbiddenEL[N][iv][\textlambda][#1]}

\newcommandx{\CIVL}[1][1=1550]{\permittedEL[C][iv][\textlambda][#1]}

\newcommandx{\HeIIL}[1][1=1640]{\permittedEL[He][ii][\textlambda][#1]}

\newcommandx{\semiOIIIL}[1][1=1666]{\semiforbiddenEL[O][iii][\textlambda][#1]}

\newcommandx{\NIIIL}[1][1=1750]{\semiforbiddenEL[N][iii][\textlambda][#1]}

\newcommandx{\CIII}{\semiforbiddenEL[C][iii]}
\newcommandx{\CIIIL}[1][1=1909]{\semiforbiddenEL[C][iii][\textlambda][#1]}

\newcommandx{\NeIVL}[1][1=2424]{\forbiddenEL[Ne][iv][\textlambda][#1]}

\newcommandx{\MgIIL}[1][1=2803]{\permittedEL[Mg][ii][\textlambda][#1]}

\newcommandx{\NeVL}[1][1=3426]{\forbiddenEL[Ne][v][\textlambda][#1]}

\newcommand{\OII}{\forbiddenEL[O][ii]}
\newcommandx{\OIIL}[1][1=3727]{\forbiddenEL[O][ii][\textlambda][#1]}
\newcommand{\OIIall}{\forbiddenEL[O][ii][\textlambda][\textlambda][3726,][3729]}

\newcommand{\NeIII}{\forbiddenEL[Ne][iii]}
\newcommandx{\NeIIIL}[1][1=3869]{\forbiddenEL[Ne][iii][\textlambda][#1]}

\newcommand{\OIII}{\forbiddenEL[O][iii]}
\newcommandx{\OIIIL}[1][1=5007]{\forbiddenEL[O][iii][\textlambda][#1]}
\newcommand{\OIIIall}{\forbiddenEL[O][iii][\textlambda][\textlambda][4959,][5007]}

\newcommandx{\NIL}[1][1=5200]{\forbiddenEL[N][i][\textlambda][#1]}

\newcommandx{\OIL}[1][1=6300]{\forbiddenEL[O][i][\textlambda][#1]}

\newcommandx{\HeIL}[1][1=5875]{\permittedEL[He][i][\textlambda][#1]}

\newcommand{\NII}{\forbiddenEL[N][ii]}
\newcommandx{\NIIL}[1][1=6584]{\forbiddenEL[N][ii][\textlambda][#1]}

\newcommandx{\OIIAuL}[1][1=7325]{\forbiddenEL[O][ii][\textlambda][#1]}

\newcommandx{\CIIFIRL}{\forbiddenEL[C][ii][\textlambda][158\,\mum]}

\usepackage[T1]{fontenc}

\DeclareRobustCommand{\VAN}[3]{#2}
\let\VANthebibliography\thebibliography
\def\thebibliography{\DeclareRobustCommand{\VAN}[3]{##3}\VANthebibliography}

\usepackage{graphicx}	
\usepackage{amsmath}	

\title[GA-NIFS: the gas and stellar properties in MACS1149-JD1 at z=9.11 with JWST]{GA-NIFS: the interplay between merger, star formation and chemical enrichment in MACS1149-JD1 at z=9.11 with JWST/NIRSpec}

\author[C. Marconcini]{C. Marconcini,$^{1,2,3}$\thanks{E-mail: cosimo.marconcini@unifi.it}
F. D'Eugenio,$^{1,4}$
R. Maiolino,$^{1,4,5}$
S. Arribas,$^{6}$
A. Bunker,$^{7}$
S. Carniani,$^{8}$
S. Charlot,$^{9}$
\newauthor
M. Perna,$^{6}$
B. Rodr\'iguez Del Pino,$^{6}$
H. \"Ubler,$^{1,4}$
C. J. Willott,$^{10}$
T. B\"oker,$^{11}$
G. Cresci,$^{3}$
M. Curti,$^{12}$
\newauthor
G. C. Jones,$^{7}$
I. Lamperti,$^{2,3}$
E. Parlanti$^{8}$
and G. Venturi,$^{8}$
\\
$^{1}$Kavli Institute for Cosmology, University of Cambridge, Madingley Road, Cambridge CB3 0HE, UK\\
$^{2}$Dipartimento di Fisica e Astronomia, Università degli Studi di Firenze, Via G. Sansone 1,I-50019, Sesto Fiorentino, Firenze, Italy\\
$^{3}$INAF - Osservatorio Astrofisico di Arcetri, Largo E. Fermi 5, I-50125, Firenze, Italy\\
$^{4}$Cavendish Laboratory, University of Cambridge, 19 J. J. Thomson Ave., Cambridge CB3 0HE, UK\\
$^{5}$Department of Physics and Astronomy, University College London, Gower Street, London WC1E 6BT, UK\\
$^{6}$Centro de Astrobiolog\'{\i}a (CAB), CSIC-INTA, Ctra. de Ajalvir km 4, Torrej\'on de Ardoz, E-28850, Madrid, Spain\\
$^{7}$University of Oxford, Department of Physics, Denys Wilkinson Building, Keble Road, Oxford OX13RH, United Kingdom\\
$^{8}$Scuola Normale Superiore, Piazza dei Cavalieri 7, I-56126 Pisa, Italy\\
$^{9}$Sorbonne Universit\'e, CNRS, UMR 7095, Institut d'Astrophysique de Paris, 98 bis bd Arago, 75014 Paris, France\\
$^{10}$NRC Herzberg, 5071 West Saanich Rd, Victoria, BC V9E 2E7, Canada
$^{11}$European Space Agency, c/o STScI, 3700 San Martin Drive, Baltimore, MD 21218, USA\\
$^{12}$European Southern Observatory, Karl-Schwarzschild-Strasse 2, 85748 Garching, Germany
}

\date{Accepted XXX. Received YYY; in original form ZZZ}

\pubyear{\the\year{}}

\begin{document}
\label{firstpage}
\pagerange{\pageref{firstpage}--\pageref{lastpage}}
\maketitle

\begin{abstract}
We present JWST/NIRSpec integral-field spectroscopy observations of the z $\sim$ 9.11 lensed galaxy MACS1149-JD1, as part of the GA-NIFS programme. The data was obtained with both the G395H grating (R$\sim$ 2700) and the prism (R$\sim$ 100). This target shows a main elongated UV-bright clump and a secondary component detected in continuum emission at a projected distance of 2 kpc.
The R2700 data trace the ionised-gas morpho-kinematics in between the two components, showing an elongated emission mainly traced by \OIIIL. We spatially resolve \OIIall, \OIIIall, and \OIIIL[4363], which enable us to map the electron density (n$_{\rm e} \sim 1.0 \times 10^3$ cm$^{-3}$), temperature (T$_{\rm e} \sim 1.6 \times 10^4$ K), and  direct-method gas-phase metallicity (-1.2 to -0.7~dex solar). A spatially resolved full-spectrum modelling of the prism indicates a north-south gas metallicity and stellar age gradient between the two components. We found 3-$\sigma$ evidence of a spatially resolved anti-correlation of the gas-phase metallicity and the star formation rate density, which is likely driven by gas inflows, enhancing the star formation in JD1. We employ high-z sensitive diagnostic diagrams to rule out the presence of a strong AGN in the main component. These findings show the unambiguous presence of two distinct stellar populations, with the majority of the mass ascribed to an old star formation burst, as suggested by previous works. We disfavour the possibility of a rotating-disc nature for MACS1149-JD1; we favour a merger event that has led to a recent burst of star formation in two separate regions, as supported by high values of \OIIIL/\Hbeta, ionised gas velocity dispersion, and gas-phase metallicity.
\end{abstract}




\section{Introduction}\label{Sec_intro}
James Webb Space Telescope (\JWST) observations of high-redshift galaxies are providing unique constraints on the galaxy and interstellar medium (ISM) properties in the very early Universe, up to z $\sim$ 10 \citep{Harikane2023, Curti2023, Larson2023, Isobe2023, Hsiao2023, Robertson2023, Curtislake2023, Sanders2024, Vikaeus2024, Carniani2024, Abdurrouf2024}. At such early epochs, \JWST reveals that galaxies are fainter and smaller compared to the local Universe \citep{Suess2022, Costantin2023, Huertascompany2024}.
As a consequence, studies aided by gravitational lensing by massive galaxy clusters are crucial to magnify the intrinsic light and sizes of distant, early galaxies, opening a new window on the investigation of their properties \citep{Vanzella2017,Vanzella2022, Mestric2022, Mowla2022, Claeyssens2023, Hsiao2023, Hsiao2023b}.
In the first few billion years of cosmic history galaxies are observed to form giant star-forming clumps, typically resulting in more irregular shapes with respect to present day analogues \citep[][]{DelgadoSerrano2010, lefevre2020, Schreiber2015, meng2021, ForsterSchreiber2011, Ferreira2020}. Observations and simulations show that gas accretion via streams onto massive clumps and during mergers triggers high rates of star formation (SF) and is the preferential driver of galaxy evolution, shaping galaxy morphology and kinematics \citep[][]{Kaviraj2014, Zanella2015, Jackson2022, BoylanKolchin2023, Zhang2023, Husko2023}. Additionally, it is crucial to characterise other drivers of galaxy evolution such as the SF activity, the gas-phase distribution, and the multi-phase gas kinematics to constrain how local galaxies built up at early epochs. 
The \JWST finally allows the exploration of these aspects with superb sensitivity and high angular resolution in the Near-Infrared (NIR).

The goal of this paper is to discuss the ionised gas and stellar population properties in the lensed galaxy MACS1149-JD1 at z = 9.11 employing \JWST/NIRSpec Integral Field Spectroscopy (IFS) observations at high and low spectral resolution. JD1 is highly magnified \citep[$\mu$ = 10;][]{Hashimoto2018, Stiavelli2023, Bradac2024} by the foreground cluster of galaxies MACS1149 at z = 0.544 \citep{Zheng2012}. This high-redshift source was first detected through a photometric excess at 4.5 $\micron$ with \Spitzer/IRAC \citep{Zheng2012}, and it was later confirmed spectroscopically with ALMA \citep[Atacama Large Millimeter Array;][]{Hashimoto2018, Tokuoka2022}. \citet{Hashimoto2018} presented tentative evidence of \Lyalpha emission via VLT/X-shooter observations and \OIII[88\micron] emission from ALMA coincident with the brighter component. To explain the observed Spectral Energy Distribution (SED) and photometric excess, they suggested the presence of a Balmer break due to a dominant old stellar component (formation redshift z $\sim$ 15) and of a younger component.
From SED fitting, \cite{Laporte2021} estimated a best-fit stellar population of age 512 Myr. Moreover, they inferred past star-formation histories (SFH) and showed that more than 93~percent of the observed stellar mass had already formed at z $\sim$ 10. \cite{Tokuoka2022} performed a detailed morpho-kinematic analysis of the ALMA data, discarding the presence of multiple clumps and suggesting the presence of a smoothly rotating gas disk. Consistent with \cite{Hashimoto2018}, \cite{Tokuoka2022} found that a $\sim$ 300 Myr old stellar population of 2.3 $\times$ 10$^{9}$ \msun is well suited to reproduce the observed features.

The advent of \JWST enabled us to drastically revise the properties of this object. \citet{Stiavelli2023} employed \JWST NIRCam and NIRSpec MSA observations to first reveal that JD1 is made up of three spatially distinct components; one in the North (hereafter, JD1-N), and two in the South (hereafter collectively JD1-S). They also showed that JD1 is mostly dust free, with a robust metal enrichment (12+log(O/H) = 7.90), and shows at most only a very weak Balmer break.
\citet{Bradac2024} presented JWST NIRCam and NIRISS observations of the system; in addition to the three unresolved clumps, they infer the presence of an underlying, edge-on disc component, containing the bulk of the stellar mass and made up of an older stellar population compared to the clumps. Taking advantage of JWST/MIRI MRS observations, \cite{marquez2023} traced ongoing SF via H$\alpha$ map, with a star formation rate (SFR) of 3.2 -- 5.3 \msun yr$^{-1}$. Moreover, they identified two main clumps with indication of a recent stellar burst ($\ge$ 5 Myr) coincident with JD1-N and a more distributed star-forming region in JD1-S.

To date, JD1 is among the highest-redshift sources for which it is possible to carry out a detailed analysis of the ISM properties via oxygen auroral-lines. In particular, thanks to the NIRSpec spectral coverage, including among many others the \OIIIL,  \OIIIL[4363] and \OIIall emission lines, we can characterise both the ionised gas kinematics and ISM properties such as the electron density and temperature \citep[][]{Kewley2019, Morishita2024, Abdurrouf2024, Mazzolari2024}.

In this work, we present spatially resolved gas properties in JD1, using new \JWST/NIRSpec Integral Field Unit (IFU) high-resolution grating (R $\sim$ 2700) and low-resolution prism (R $\sim$ 100) observations. This paper is organised as follows. In Sec. \ref{Sec2_observations} we present the NIRSpec observations, together with the data reduction. Section \ref{Sec3_analysis_line_fitting_line_ratios} is devoted to the description of emission line analysis and the following results on the ISM properties and gas kinematic. In Sec. \ref{Sec4_discussion} we discuss the impact of our results in the broad context of galaxy assembling processes at high-redshift. Finally, in Sec. \ref{Sec_conclusions} we summarise the main results. In this paper, we adopt a $\Lambda$CDM cosmology: H$_0$: 67.4 km s$^{-1}$ Mpc$^{-1}$, $\Omega_\mathrm{m}$ = 0.315, and $\Omega_\Lambda$ = 0.685.


\section{Observations and data reduction}\label{Sec2_observations}

\subsection{NIRSpec IFU Observations}\label{Sec2_nirspec_observations}
MACS1149-JD1 (hereafter JD1, R.A.  11$\rm ^h$ 49$^m$ 33.580$^s$ DEC +22$^{\circ}$ 24\arcmin 45.70\arcsec, J2000) was observed as part of the Galaxy Assembly with NIRSpec Integral Field Spectroscopy (GA-NIFS\footnote{GA-NIFS website: \url{https://ga-nifs.github.io}}, PIs: S. Arribas, R. Maiolino) Guaranteed Time Observations (GTO) programme, included in Program ID $\#$1262 (PI: N. L\"utzgendorf). The observations were carried out on June 5 2023, using the NIRSpec IFS \citep{Jakobsen2022,Boker2022}, covering a 3.1\arcsec $\times$ 3.2\arcsec region, with a native spaxel size of 0.1\arcsec \citep[][]{Boker2022, Rigby2023}. The observations consist of two configurations: high spectral resolution with the G395H grating (R$\sim$ 2700, covering 2.87--5.14~\micron) and low spectral resolution with the prism (R$\sim$100, covering 0.6--5.3~\micron). Both observations used a medium (i.e. $\sim$ 0.5\arcsec) cycling pattern of eight dithers. For G395H we used 31 groups per integration and one integration per exposure, with the NRSIRS2 readout mode \citep{Rauscher2017}, giving a total integration time of 18207 seconds (5.05~h).
For the prism we used 33 groups per integration and one integration per exposure with the NRSIRS2RAPID readout, giving a total integration time of 3968 seconds (1.1~h). 

JD1 resides behind the MACS J1149.6+2223 cluster and was observed by ground- and space-based instruments providing insightful information on the multi-band properties of this system as discussed in Sec. \ref{Sec_intro}. Similarly to the magnification analysis performed in \citet{Bradac2024}, we estimated the 2D spatial magnification using different publicly available lens models. Overall, all models show negligible spatial variation of the magnification factor $\mu$ over the spatial scale of JD1, despite on average different lens models can provide very different different median magnification (see \citet{Bradac2024}). Therefore, in this work we adopted a $\mu$ = 10 $\pm$0.7 and corrected for the lensing magnification all the appropriate properties, thus assuming that all flux-dependent quantities are in units of (10/$\mu$) \citep{Hashimoto2018, marquez2023, Stiavelli2023, Bradac2024}.

\begin{figure*}
	\includegraphics[width=\linewidth]{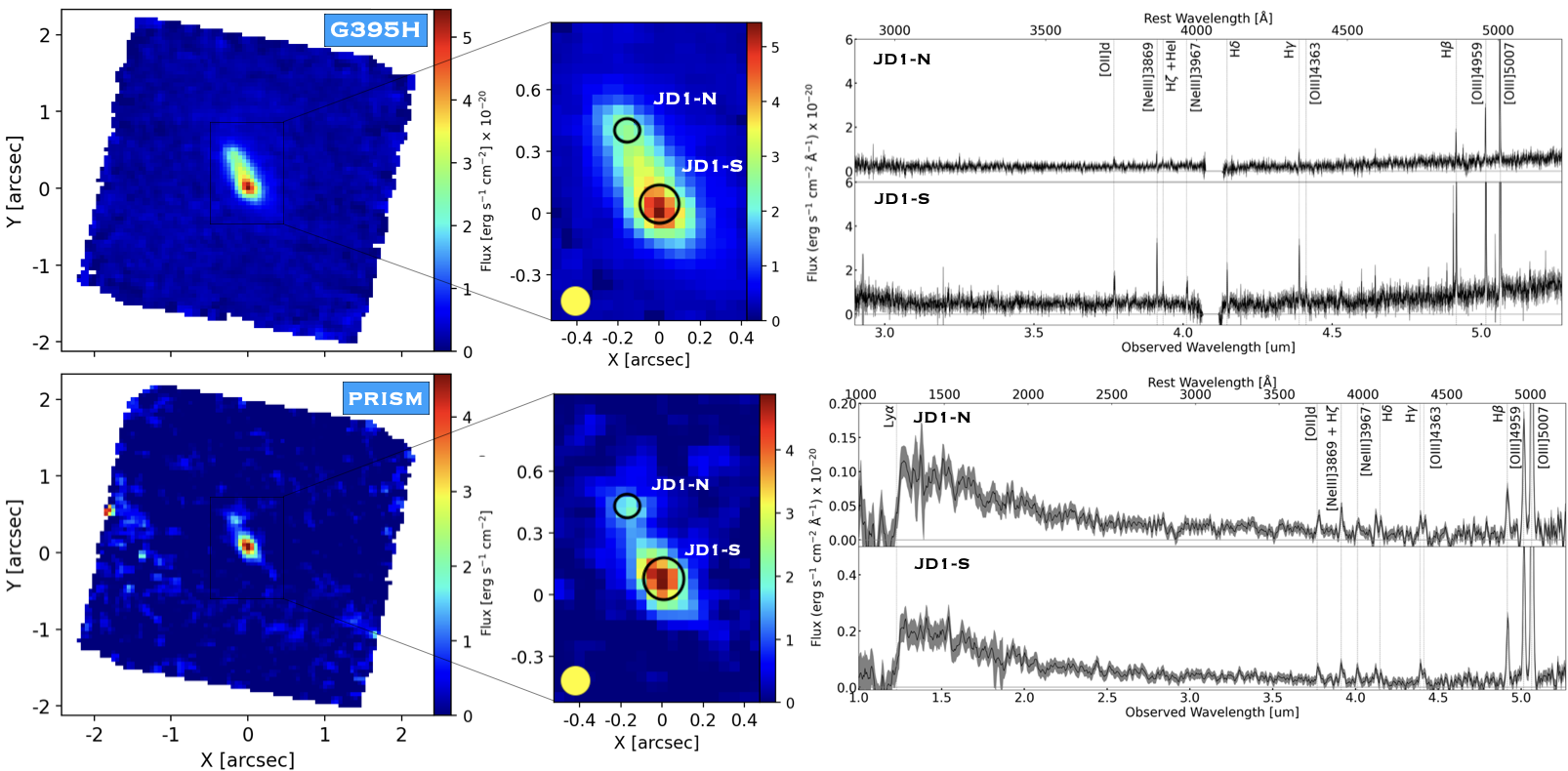}
    \caption{JWST NIRSpec collapsed images and spectra before continuum subtraction. Left: G395H (top) and prism (bottom) data cubes collapsed over the total \OIIIL emission line and 1.75 - 2.23 $\micron$, respectively. Middle: Zoom-in centred on JD1 of the data cubes shown in the left panels. Yellow filled circles represent the fiducial PSF size. Right: Integrated spectra from JD1-N and JD1-S for the G395H (top) and prism (bottom) data cubes extracted from the apertures shown in middle panels. Labels indicate the identified emission lines. North is up and East is to the left.}
    \label{fig:grating_prism_map}
\end{figure*}

\subsection{Data Reduction}\label{Sec2_data_reduction}
We performed the data reduction using the well tested \JWST calibration pipeline version 1.8.2 with CRDS context jwst1068.pmap. The steps and changes we performed with respect to the standard pipeline in order to improve the final data quality are discussed in \cite{perna2023_pipeline}. To combine each integration and create the final data cube with spaxel size of 0.05\arcsec, we adopted the $drizzle$ method, with particular attention to use an official patch to account for a known issue affecting the calibration pipeline\footnote{\url{https://github.com/spacetelescope/jwst/pull/7306}}. For the G395H data we performed no background subtraction as we will account for its spectral contribution during the emission line fitting.
For the low-resolution prism instead, we performed the background subtraction on the detector image, for each of the eight observations separately.
From a first datacube, we created a source mask including all contiguous regions with detected \OIIIL; we enlarged this mask by adding padding of width 0.1 arcsec.
We then deprojected this on-sky source mask back onto each of the eight 2-d detector images, using the function \textsc{blot} in the \textsc{jwst} pipeline v1.12.5.
For each 2-d `cal' image produced by the pipeline stage 2, we fit a simple linear slope across each of the slices and at each pixel along the dispersion direction, using the 2-d mask to exclude pixels where the source is present. We then smooth the resulting background in the dispersion direction using 7-pixel median filtering.
The resulting background images are subtracted from the cal images.
We then re-run the pipeline stage 3 on the background-subtracted 2-d images.

\section{Analysis and Results}\label{Sec3_analysis_line_fitting_line_ratios}
\subsection{Emission line fitting}\label{Subsec3_line_fitting}
The G395H high-resolution data cube was analysed by means of a set of custom Python scripts with the goal of subtracting the continuum and then fit the observed emission lines \citep[e.g. see][]{Marasco2020, tozzi2021}. First, we performed Voronoi binning \citep{Cappellari2003} on the continuum level requiring  a minimum S/N in each spectral channel of 5. Then we used the Penalized Pixel-Fitting software \citep[\textsc{pPXF};][]{Cappellari2004, Cappellari2023} to fit the continuum in each bin
with a second order polynomial while simultaneously fitting the emission lines with one Gaussian component. During this procedure, we masked bad pixels throughout all the data cubes. Then, we subtracted the continuum spaxel by spaxel and matched the spatial resolution within the Field of View (FoV) by convolving each spectral slice of the data cube with a $\sigma$ = 1 pixel ( i.e., 0.05 \arcsec) Gaussian kernel.
Finally, we fit the emission lines of the continuum-subtracted cube with one or two Gaussians while imposing the same velocity and widths of
each component for all the lines and leaving the intensities as a free parameter. For the emission line doublet \OIIIall we fixed an intrinsic ratio of 1/3 between the two lines \cite{Osterbrock2006}. To decide the number of Gaussian components in each spaxel we performed a reduced $\chi^2$ analysis with the purpose of identifying any potential region where two components might be needed. In particular, we compared the residuals of the fitting in the wavelength range of the \OIIIall doublet using a Kolgomorov-Smirnov test to check if the residuals with an additional Gaussian component are statistically different from those of a model with a single component \citep[see][for details]{Marasco2020}. As a result, we adopted two Gaussian components only in a few spaxels where a broad line wing is detected. Finally, we obtained an emission-line model cube for each emission line. 

\begin{figure*}
	\includegraphics[width=\linewidth]{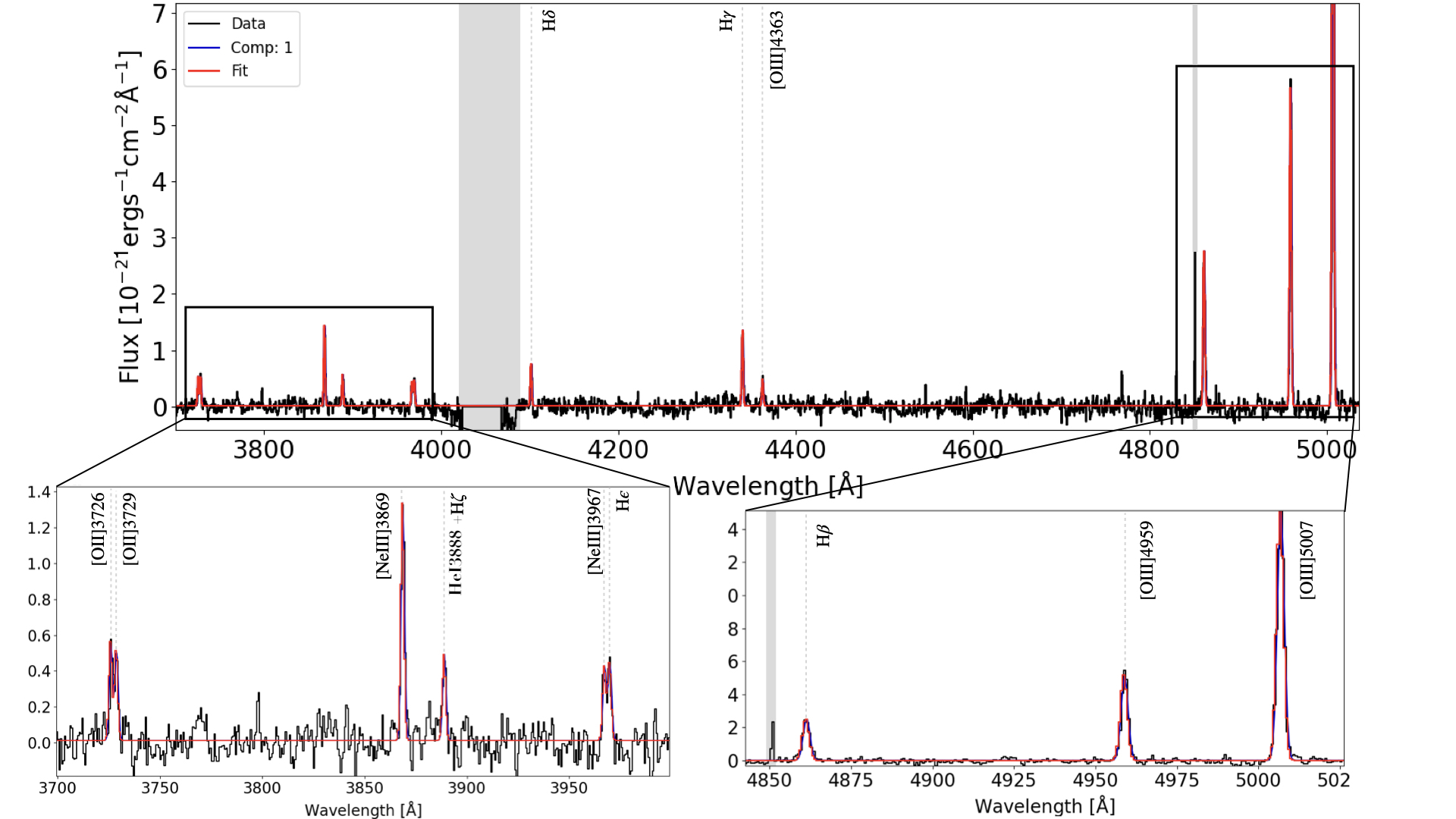}
    \caption{JWST/NIRSpec G395H spectrum extracted from a single spaxel of JD1-S. The top panel shows the total spectrum focusing on the 3700 - 5050 \AA \ spectral region, which contains the emission lines of interest. The solid black and red lines show the observed spectrum and best-fit model, respectively. The grey dashed lines show the location of the fitted emission lines. The bottom two panels show a zoom on the regions containing the \OII and \NeIII doublets, \Hepsilon, \Hzeta+HeI$\lambda$3888 (left), and the \Hbeta and \OIII complex (right). Shaded grey areas are masked spectral channels. }
    \label{fig:spectrum_fit_ex}
\end{figure*}

The left panels in Fig. ~\ref{fig:grating_prism_map} show images of the G395H data cube obtained collapsing over the total \OIIIL emission and the prism data cube collapsed in the wavelength range 1.75 - 2.23 $\micron$ (i.e. 1700 - 2200 \AA rest-frame), corresponding to F200W band in NIRCam. While the G395H image is dominated  by the nebular emission, the prism one mainly traces the continuum emission with somewhat better angular resolution due to the sharper \JWST PSF at these shorter wavelengths. The right panels in Fig.~\ref{fig:grating_prism_map} show the integrated spectra of the high- and low-resolution (background subtracted) data cubes extracted from the apertures marked with black circles in the central panels. North and south apertures have radii of 0.1\arcsec (i.e. $\sim$ 450 pc) and 0.15\arcsec (i.e. $\sim$ 670 pc), respectively. The two apertures correspond to the putative location of the two components identified by \citet{Hashimoto2018} (see also \citealp{Stiavelli2023}).
Besides the known components of JD1, we did not detect any other source in the 3-by-3-arcsec field of view of NIRSpec/IFS.
Therefore, in this work, we focused on a subcube centred on the main galaxy and shown in the left and middle columns of Fig.~\ref{fig:grating_prism_map}. Overall, from the high resolution data cube we fitted the following emission lines: \Hbeta, \Hgamma, \Hdelta, \Hepsilon, \Hzeta, \OIIL[3726], \OIIL[3729], \HeIL[3889], \NeIIIL, \NeIIIL[3967],  \OIIIL[4363], \OIIIL[4959], and \OIIIL. The NIRSpec G395H spectrum extracted from a pixel in JD1-S is shown in Fig.~\ref{fig:spectrum_fit_ex} with the single Gaussian model used to fit the aforementioned emission lines.
\subsection{Gas kinematics}\label{Subsec3_kinematic}
\begin{figure}
	\includegraphics[width=\linewidth]{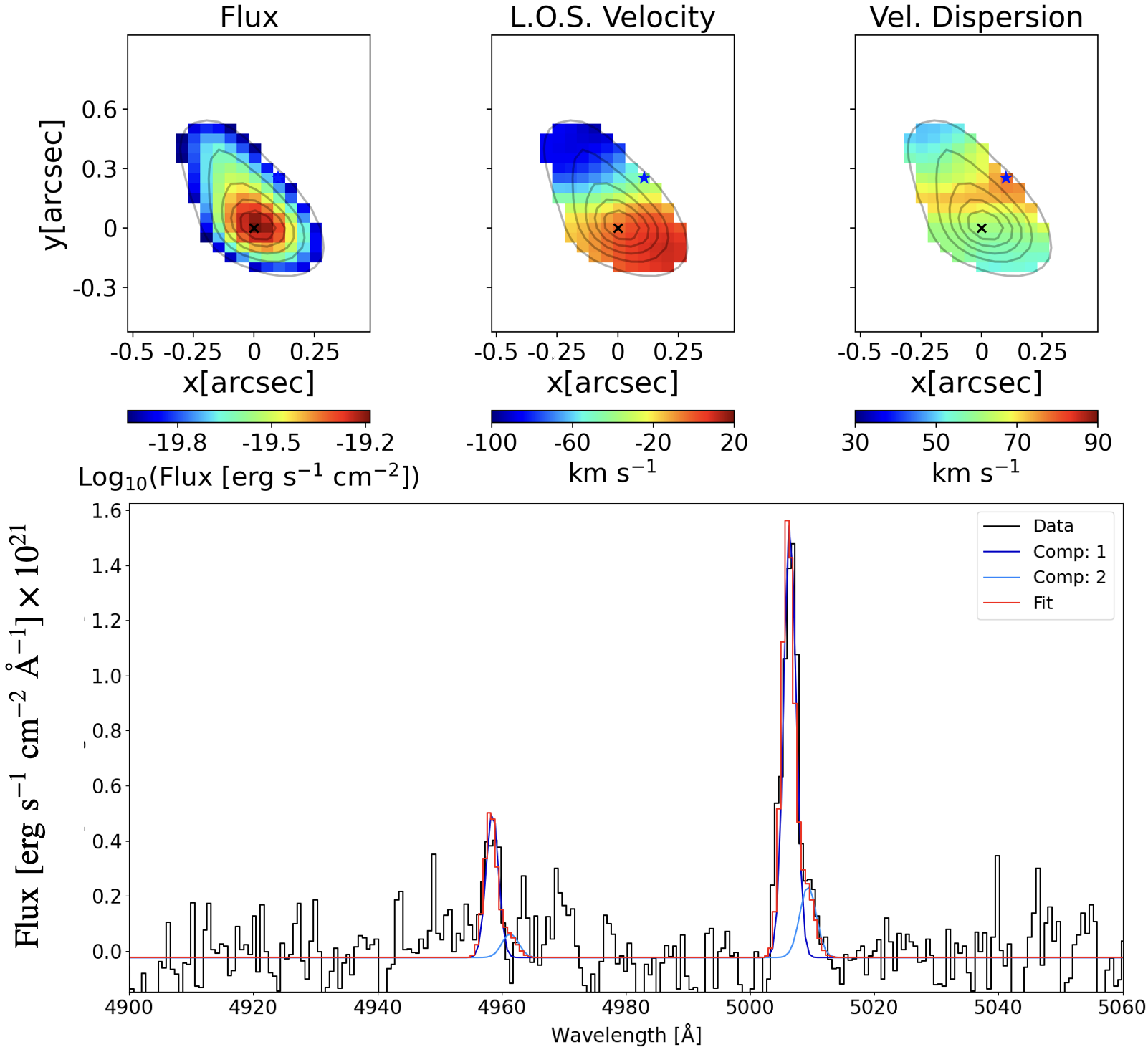}
    \caption{\OIIIL G395H moment maps and zoom of the \OIII complex. Top panel: From left to right the integrated emission line flux, the flux-weighted LOS velocity and the velocity dispersion maps. All moment maps are calculated from the total best fit model. Moment maps are masked at S/N smaller than 3. Black contours are arbitrary \OIIIL levels. Bottom panel: Zoom of the \OIIIall emission line doublet extracted from the spaxel marked with the blue star. Data and best fit model are in black and red, respectively. The two Gaussian components are represented in blue and light blue solid lines.}
    \label{fig:OIII007_mom_maps}
\end{figure}

As a result of the spaxel-by-spaxel fit described in the previous section we obtain the one- or two-Gaussian best-fit model in each spaxel. From these models, we create moment maps for all emission lines of interest, which provide precious information on the projected gas kinematics. By comparing these maps and as a consequence of the fit kinematic constraints described in the previous section, we conclude that there is no statistically significant difference between the kinematics of different emission lines. For this reason, in the following we focus only on the \OIIIL emission line, which has the highest S/N. Fig. \ref{fig:OIII007_mom_maps} shows the emission-line flux (moment 0), the flux-weighted line of sight (L.O.S.) velocity (moment 1), and the velocity dispersion maps (moment 2) for the \OIIIL emission line, computed from the best-fit emission line model cube. The zero velocity is set assuming z = 9.1130 as a reference. Black contours show arbitrary \OIIIL flux levels. The peak of the \OIIIL emission line originates from the main UV clump (JD1-S), with a blue-shifted tail elongated towards JD1-N. The right panel shows a horizontal central region with high velocity dispersion, which may be due to the beam smearing of the strong velocity gradient. However, the map shows also an intriguing excess of velocity dispersion towards the Western region. The bottom panel of Fig. \ref{fig:OIII007_mom_maps} shows the spectrum extracted from a spaxel in such region of enhanced velocity dispersion, with the best-fit multi-Gaussian model in red. We found that over this region of high velocity dispersion the fits benefit from a second Gaussian component to reproduce the observed \OIIIL emission line profile based on $\chi^2$ minimisation, likely indicating the presence of outflow or merger. Similarly, \citet{marquez2023} analysed MIRI/MRS observations tracing the H$\alpha$ kinematics and measured a difference in velocity dispersion of $\approx$ 60 km s$^{-1}$ among the two clumps, supporting the hypothesis of outflowing gas powered by the UV-bright clump JD1-S \citep[][see also Sec. \ref{Subsec_discussion_rotating_nature}]{Tokuoka2022}.

\begin{figure}
	\includegraphics[width=\linewidth]{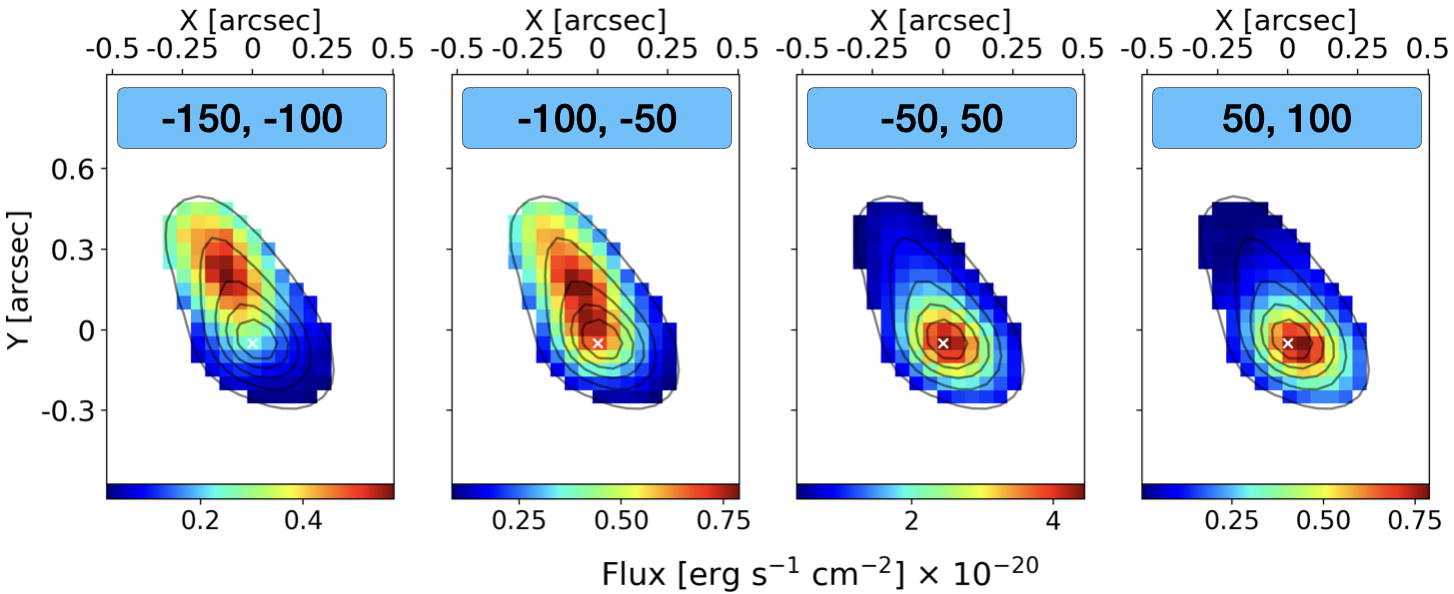}
    \caption{\OIIIL collapsed images at different velocity channels, assuming as zero-velocity the redshift z = 9.1130. From left to right the intervals are: (-150,-100), (-100,-50), (-50,50), (50,100) km s$^{-1}$. }
    \label{fig:OIII5007_velchan_maps}
\end{figure}

From the peak of integrated \OIIIL emission line over the two components we measured redshifts of z = 9.1130 $\pm$ 0.0002 and z = 9.1099 $\pm$ 0.0002, for JD1-S and JD1-N, respectively. Therefore, assuming the redshift of JD1-S as reference, JD1-N is blue-shifted by $\sim$ 90 km s$^{-1}$. The zero-order moment map in Fig. \ref{fig:OIII007_mom_maps} shows the \OIIIL emission of ionised gas connecting the two components. The channel maps of \OIIIL in Fig. \ref{fig:OIII5007_velchan_maps} highlight this blue-shifted elongated tail, showing a clear velocity gradient from the JD1-N to JD1-S, with average projected velocities between -150 and 0 km s$^{-1}$.

From these kinematic maps alone, we cannot conclusively distinguish between a rotating disc scenario or a merger. Indeed, in both cases we would expect to see the velocity dispersion peaking halfway between the blue-shifted and red-shifted components (Fig.~\ref{fig:OIII007_mom_maps}). The kinematic axis of the putative disc seems tilted by $\approx 10^\circ$ relative to the major axis of the photometry \citep[which runs approximately along the line connecting JD1-N to JD1-S;][]{Bradac2024}. The possible disc or merger nature of JD1 will be discussed more in-depth in Sec.~\ref{Subsec_discussion_rotating_nature}.

From the spatially resolved \OIIIL emission-line fit we measured an average velocity dispersion of 62 $\pm$ 7 km s$^{-1}$, consistent with ALMA \citep{Hashimoto2018, Tokuoka2022}.
To calculate a reference dynamical mass, we assumed the structure of JD1 to be consistent with a disc and used the stellar virial estimator of \citet{vanderWel2022}, using our velocity dispersion, a (de-lensed) effective radius of 585$^{+157}_{-52}$ $\sqrt{10/\mu}$ pc and S\'ersic index of 1 $\pm$ 0.2 measured by \citet{Bradac2024} from NIRCam observations, and an axis ratio of 0.6. We obtained a dynamical mass of M$_{\rm dyn}$ = 1.2$^{+0.5}_{-0.4}$ $\sqrt{10/\mu}$ $\times$ 10$^{9}$\msun. This value is consistent with previous \Halpha-based estimates of 2.4 $\pm$ 0.5 $\times$ 10$^{9}$~\msun from \cite{marquez2023} and with resolved \OIIIL[88\micron]-based estimates of 0.7 - 3.7 $\times$ 10$^{9}$\msun of \cite{Tokuoka2022}.
Using different virial calibrations \citep[e.g.,][]{cappellari+2006,cappellari+2013a, wisnioski+2018} would give dynamical masses in the range $0.8\text{--}2.6\sqrt{10/\mu} \times 10^{9}$\msun, which does not change our conclusions.


\subsection{Excitation Diagnostics and Line Ratios}\label{Subsubsec3_diagnostic}
\begin{figure}
	\includegraphics[width=\linewidth]{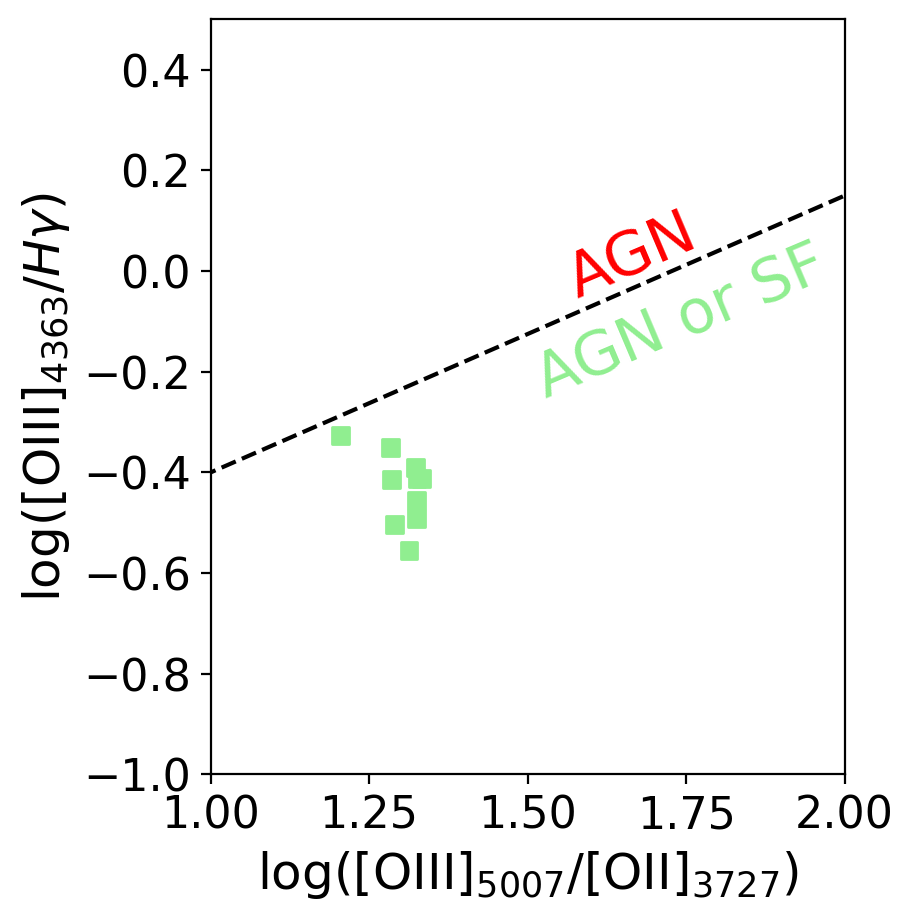}
    \caption{AGN-diagnostic diagram of \citet{Mazzolari2024} for the excitation source in JD1 based on the \OIIIL[4363]/\Hgamma vs \OIIIL/\OIIL[3727] line ratio. Dashed black line is the demarcation between pure AGN (above) and AGN or SF excitation source (below).}
    \label{fig:diagnostic_diagram}
\end{figure}

We used the results of the emission line fitting of the high-resolution data cube to examine the excitation source in JD1.
Recent works have highlighted the difficulty in separating SF and AGN ionisation at high z and low metallicity by using the standard BPT diagram \citep[][]{Feltre2016, nakajima2022, Kocevski2023, Ubler2023, Scholtz2023, Maiolino2023AGN_JADES}.
We therefore explored the possible source of ionisation by employing the new diagnostic diagrams proposed by \cite{Mazzolari2024}, which use the ratio of \OIIIL[4363]/\Hgamma to identify narrow-line Type II AGN at high redshift. 
Fig.~\ref{fig:diagnostic_diagram} shows one of such diagnostic diagrams using the \OIIIL[4363]/\Hgamma vs \OIIIL/\OIIIL[4363]  line ratios, together with a map of the most likely excitation source in JD1 and the \OIIIL[4363]/\Hgamma line ratio. In this diagram, being above the demarcation line is a sufficient but not necessary condition for an object to be identified as an AGN, therefore spaxels which are below the demarcation line can still be potentially associated with AGN excitation. The diagram shows that no regions in JD1 show unambiguous evidence for AGN. Beside the one shown in Fig. \ref{fig:diagnostic_diagram} we also explored the other two diagrams discussed in \citet{Mazzolari2024} (employing \OIIIL/\OIIIL[4363] and \NeIIIL[3869]/\OIIall emission line ratio vs \OIIIL[4363]/\Hgamma) and observed that all the selected spaxels in each diagram do not show unambiguous evidence for an AGN, favouring the SF ionisation as most likely scenario. For completeness, we estimated the position of JD1 in the standard BPT diagram adopting an \Halpha flux of 1.05 $\pm$ 0.07 $\times$ 10$^{-17}$ erg s$^{-1}$ cm$^{-2}$, a 3$\sigma$ upper limit flux on \NII of 2.1 $\times$ 10$^{-18}$ erg s$^{-1}$ cm$^{-2}$ from \citet{marquez2023} and an average log(\OIII/\Hbeta) = 0.8 ratio from our spatially resolved analysis. Similarly to the result obtained with the diagnostic diagram shown in Fig. ~\ref{fig:diagnostic_diagram}, the standard BPT diagram report an ambiguous ionisation nature for JD1, with the location of this system being located on the demarcation line between SF and AGN ionisation. Therefore, from the results of both diagnostics we cannot exclusively state the nature of the main ionisation source in JD1.

\begin{figure}
	\includegraphics[width=\linewidth]{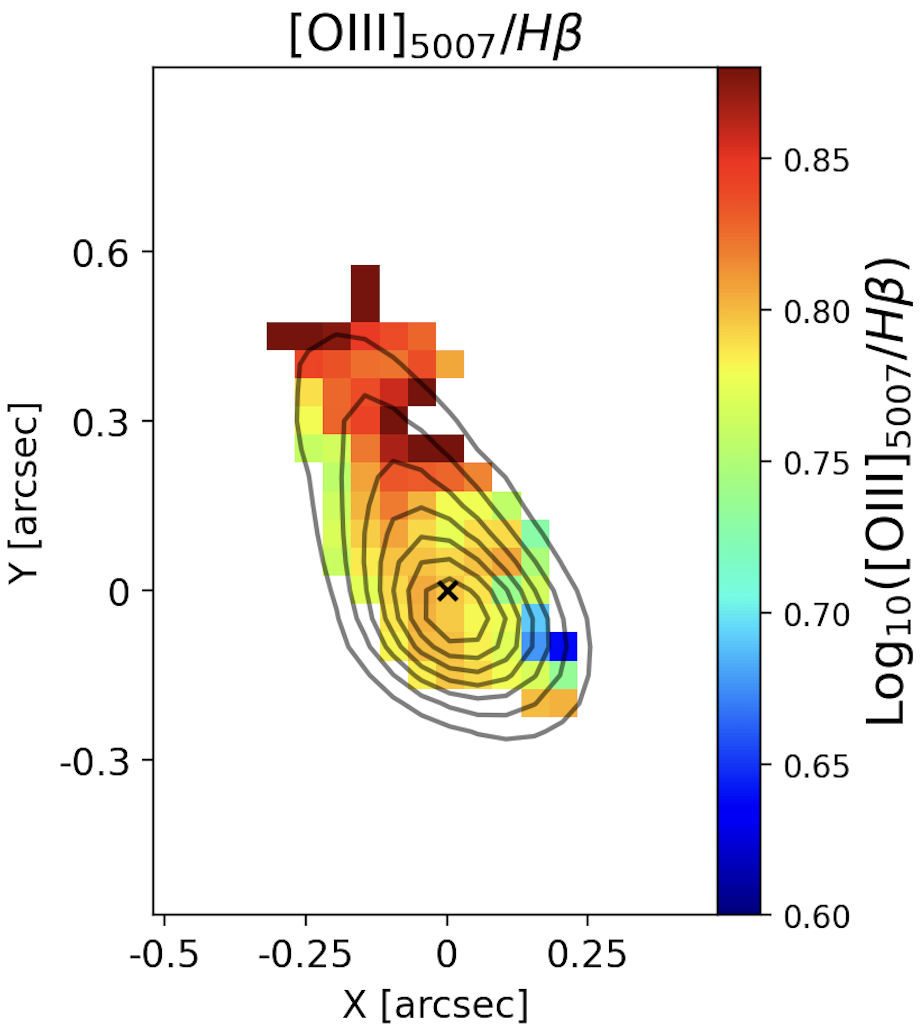}
    \caption{\OIIIL/\Hbeta emission line ratio in JD1 in logarithmic scale. Black solid line are arbitrary \OIIIL levels. Spaxels with S/N $\le$ 3 are masked.}
    \label{fig:line_ratios}
\end{figure}

In order to trace regions of high ionisation we also computed the \OIIIL/\Hbeta line ratio map shown in Fig. \ref{fig:line_ratios}. Due to the difference in ionisation potential of \OIIIL and \Hbeta this ratio is a good indicator of the hardness of the radiation field, and thus of instantaneous bursts of SF. We found that JD1-N is characterised by high excitation, with \OIIIL/\Hbeta $\sim$ 8, at variance with JD1-S which is characterised by an average \OIIIL/\Hbeta $\sim$ 6. 

\subsection{Electron density and Temperature}\label{Subsub3_electron_den_temperature}
The high spectral resolution G395H data includes the spectrally resolved \OIIall doublet and \OIIIL[4343] emission lines, which allow us to compute the first spatially resolved estimate of the electron density and temperature at z$\geq$ 9.
In particular, first we used the \textsc{pyneb} Python package \citep[][]{Luridiana2012, Luridiana2015} to compute the electron temperature T$_{\rm e}$ in the \OIII-emitting region from the \OIIIL/\OIIIL[4363] line ratio assuming no dust attenuation (\citealp{Hashimoto2018}; in any case, even a dust attenuation $A_V=1$~mag would only change T$_{\rm e}$ by 20 percent) and an electron density of 600 cm$^{-3}$ (which is a good assumption based on the value measured in high-z JWST galaxies; \citealp{Isobe2023, Abdurrouf2024}).
To compute the temperature of \OII-emitting regions (T$_2$), due to the absence of temperature diagnostics, we used the relation T$_2$ = -0.744 + T$_{\rm e}$ $\times$ (2.338 - 0.610$\times$ T$_{\rm e}$) from \cite{Izotov2006} to derive it from the temperature of the \OIII-emitting regions. 

Using T$_2$ and the observed \OIIL[3726]/\OIIL[3729] ratio, we then compute the electron density.
Fig. \ref{fig:density_temperature_electrons} shows spatially resolved maps of the electron gas density (left panel) and temperature (middle panel). The derived gas-phase metallicity computed with the direct T$_\mathrm{e}$ method is discussed in the Sec. \ref{Subsec3_metallicity}, together with an estimate via prism continuum fitting. The average electron temperature and density of JD1-S are 1.4$^{+0.6}_{-0.7}$ $\times$ 10$^{4}$ K and 670$^{+100}_{-240}$ cm$^{-3}$, with the latter being consistent with the assumed density to compute T$_{\rm e}$ . As shown in Fig. \ref{fig:density_temperature_electrons}, due to the low S/N of the \OII doublet over JD1-N we could not compute a spatially resolved estimate of the electron density. Therefore, we integrated the spectrum from the aperture of JD1-N (see Fig. \ref{fig:grating_prism_map}) and assuming an electron temperature of 10$^{4}$ K we estimated an electron density from integrated spectrum of 730 cm$^{-3}$, which is slightly higher than the integrated value of JD1-S but still consistent within the uncertainties. We computed the uncertainties on the electron density on the basis of a Monte Carlo technique. Our findings are consistent with the observed redshift evolution of the electron density reported by \citet[][see also \citealp{Abdurrouf2024}]{Isobe2023}. In particular, it is possible that, due to the more compact morphology of gas nebulae and the lower metallicity at high redshift, the electron density is higher than in the local Universe (see Sec. \ref{Sec4_discussion} for a more detailed discussion).

\subsection{Continuum fitting - Stellar population properties}\label{Subsec3_continuum_fitting}
To provide a spatially resolved analysis of the stellar mass in JD1, we performed the continuum fitting of the prism data cube both spaxel by spaxel and by integrating the total spectrum using \textsc{prospector} \citep[][]{Johnson2021}, a Bayesian SED modelling framework built around the stellar-population synthesis tool \textsc{fsps} \citep{Conroy2009, Conroy2010}.
We performed a SED modelling of the observed spectrum from 0.6 to 5.27~\micron following the procedure described in \citet{Tacchella2023} and \cite{perez_gonzales2023}. Using the prism data cube we performed a spatial smoothing, to obtain approximately the same PSF at all wavelengths. We assumed a 2-d Gaussian kernel, with different full-width half-maximum values along and across slices, as a function of wavelength \citep{deugenio2023}. 

For the modelling, we configured \textsc{fsps} to use the MILES stellar atmospheres \citep{Falconbarroso2011} and MIST isochrones \citep{Choi2016}. The nebular emission is modelled using pre-computed grids from \textsc{cloudy} \citep{Ferland1998}, as described in \citet{Byler2017}; this approach takes into account possible stellar absorption at the wavelength of emission lines \citep[see e.g.][]{perez_gonzales2003, perez_gonzales2008}. Finally, we accounted for dust attenuation using a flexible dust attenuation law, consisting of a modified Calzetti law \citep[][]{Calzetti2000} with a variable power-law index and UV-bump strength \citep{Noll2009, Kriek2013}. Stars younger than 10~Myr are further attenuated by an extra dust screen, parametrised as a simple power law \citep{Charlot2000}.
No dust emission is included, because our reddest wavelengths are still in the rest-frame optical range, where dust emission is negligible. The star-formation history (SFH) uses 9 fixed time bins between z = 9.1130 and z = 20; the first three bins are at 10~Myr, 30~Myr, and 100~Myr, the remaining 6 bins are logarithmically spaced in time. We use a continuity prior to relate the log ratio of the SFRs between adjacent time bins \citep{Leja2019}.
The model free parameters and their prior probability distributions are listed in Table~\ref{t.prospector}. The left panel in Fig.~\ref{fig:stellar_mass_prospector} shows the spatially resolved stellar mass surface density estimated with this method, together with the spatially resolved maps of the SFR surface density within the last 10 (SFR$_{10}$) and 100 Myr (SFR$_{100}$).

From the integrated spectrum of JD1, we found a total stellar mass budget of $\log(\mstar/\msun)_{\rm JD1}=7.47^{+0.05}_{-0.05} \log (10/\mu)$, which is on average an order of magnitude lower compared to previous estimates of the stellar mass in JD1 \citep{Laporte2021, Stiavelli2023, Bradac2024}. Such a significant discrepancy has to be ascribed to differences in the data. Our prism spectrum displays a clearly detected continuum, with evidence for a Balmer jump due to nebular emission (Fig.\ref{fig:grating_prism_map}). In contrast, other studies find evidence of a Balmer break, surmising an old stellar population which greatly contributes to the total stellar mass. In two cases, the Balmer break is seen only from photometry (\citealp{Laporte2021} and \citealp{Bradac2024}, their fig.~2G), whereas in \citet{Stiavelli2023} the Balmer break is not clearly seen in the spectroscopy (their fig.~2). 
Also, our estimate of the total stellar mass is 0.4~dex lower than the value we obtain from modelling the spectrum in each spaxel and then adding up the resulting stellar masses (which gives $\log(\mstar/\msun)=7.88 \log(10/\mu)$). This discrepancy may be due to outshining of fainter, older components by younger stars in the integrated flux \citep{GimenezArteaga2023}.
From the spatially resolved kinematic analysis (see Sec. \ref{Subsec3_kinematic}) and assuming a rotating disc, we estimated a (de-lensed) maximal dynamical mass of
M$_{\rm dyn}$ = 1.2$^{+0.5}_{-0.4}$ $\sqrt{10/\mu}$ $\times$ 10$^{9}$\msun.
This value is much higher than the total stellar mass in JD1 and is possibly due either to a large cold gas content, which is feeding the ongoing SF \citep[e.g.][]{marquez2023}, or, alternatively, to non-virialised motions (we discuss this finding in Sec. \ref{Sec4_discussion}) or large amounts of dark matter \citep{deGraaff2024} or a combination of the aforementioned effects.

\begin{figure*}
	\includegraphics[width=\linewidth]{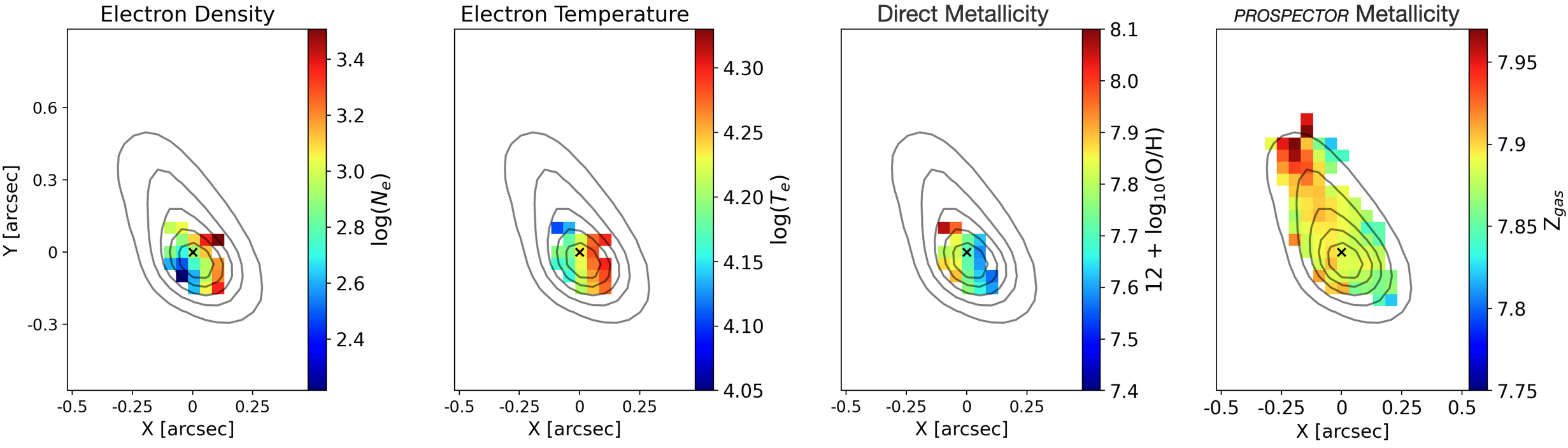}
    \caption{ISM physical properties of JD1. From left to right: Electron density, temperature, gas-phase metallicity derived with the direct method (T$_\mathrm{e}$ method) and gas-phase metallicity derived with \textsc{prospector} performing SED fitting (see Sec. \ref{Subsub3_electron_den_temperature} and Sec. \ref{Subsec3_continuum_fitting} for details). Spaxels at S/N $\le$ 3 for the \OIIall doublet are masked. Black contours are arbitrary \OIIIL flux levels. The black cross marks the position of the \OIIIL line peak.}
    \label{fig:density_temperature_electrons}
\end{figure*}

From the SED fitting of the integrated spectrum we derived a SFR$_{10} = 2.8^{+0.1}_{-0.1}$ \msun yr$^{-1}$ and SFR$_{100} = 0.29^{+0.1}_{-0.1}$ \msun yr$^{-1}$, integrating the SF history (SFH) of the last 10 and 100 Myr, respectively. Combining the result for the total stellar mass formed and the SFR, we computed the sSFR within the last 10 and 100 Myr. The resulting spatially resolved maps are shown in Fig. \ref{fig:specific_SFR}. From the integrated spectrum instead, we derived specific SFR of sSFR$_{10}$ = 9.6$^{+1.9}_{-1.7}$ Gyr$^{-1}$ and sSFR$_{100}$ = 0.97 $^{+0.2}_{-0.2}$ Gyr$^{-1}$. The sSFR maps show consistent values between JD1-N and JD1-S in the last 100 Myr. Within the last 10 Myr instead, we observe higher sSFR over JD1-S compared to JD1-N. Nonetheless, comparing the sSFR within the last 10 and 100 Myr for JD1-N, we conclude that the majority of its mass formed very recently \citep{marquez2023, Stiavelli2023, Bradac2024}, which could explain the observed enhancement of the gas-phase metallicity (in absence of significant inflows of low-metallicity gas; see next section and right panels of Fig. \ref{fig:density_temperature_electrons}).

As a comparison with the SFR derived via SED fitting, we measured the \Hbeta flux to provide a direct estimate of the SFR in JD1. We assumed a Case B recombination scenario \citep{Osterbrock2006} and based on the measured \Hbeta flux of 1.47 $\pm$ 0.2 erg s$^{-1}$ cm$^{-2}$, we computed the SFR as:
\begin{equation}
    SFR(\Hbeta) = 5.5 \times 10^{-42} \  L_{\Hbeta} (\mathrm{erg \, s^{-1}}) \times f_{\Halpha/\Hbeta} \ \msun \mathrm{yr}^{-1},
\end{equation}\label{eq.1}\noindent
where $f_{\Halpha/\Hbeta}$ = 2.85 for T$_\mathrm{e}$ = 1.4 $\times$ 10$^4$ K (see Sec. \ref{Subsub3_electron_den_temperature}. Correcting for the lensing factor we obtain $\rm  SFR(\Hbeta)$ = 2.5 $\pm$ 0.3  (10/$\mu$) $\msun \mathrm{yr}^{-1}$, which is consistent with the value derived via SED fitting.

Fig. \ref{fig:EW_oiii_hb} shows maps of estimated rest-frame equivalent width (EW) of the \Hbeta and \OIIIL emission lines. Both maps show a mild decrease of EW from the centre to JD1-S outwards. A similar trend of radially decreasing EW of \Hbeta in high-redshift galaxies has already been reported \citep{Tripodi2024}.
We also find higher EW values over JD1-S compared to JD1-N, as also shown by the sSFR maps in Fig. \ref{fig:specific_SFR}.
Both findings points towards a higher and more recent SF in JD1-S, consistent with the result obtained via SED fitting. Moreover, as discussed in Sect. \ref{Subsec_discussion_rotating_nature}, JD1-N may be characterised by an extreme escape fraction of ionising photons. The effect of an elevated escape fraction over JD1-N translates in a loss of ionising photons that in turn reduces the intensity of emission lines with respect to the continuum, decreasing the observed EW. We note that the SED fitting algorithm that we used for the continuum analysis is not tailored to account for any fraction of escaping photons. Similarly to the effect on the EW, including a higher escape fraction in JD1-N would enhance the estimated resolved SFR and sSFR density maps, consistently with a scenario of a recently formed stellar population.

Fig. \ref{fig:beta_slope} shows a map of the $\beta_{UV}$ slope, measured between 1200 to 1500 $\AA$ rest-frame (where $\beta_{UV}$ is defined as $F_\lambda \propto \lambda ^{\beta_{UV}}$), computed via continuum fitting of the prism data cube. Interestingly, we observed a smooth north--south gradient, highlighting different properties of the two main companions. The $\beta_{UV}$ has a maximum of --2.3 over JD1-S and then decreases to a steeper slope of --2.6 towards JD1-N.
Given that we find no evidence of dust from the emission-line ratios, these differences in UV slope likely indicate an older and younger stellar population for JD1-N and JD1-S, respectively \citep[][]{Stiavelli2023, Bradac2024}. At variance with previous works, we do not detect any Ly$\alpha$ emission in the low-resolution data cube, probably due to the low spectral resolution of our prism data \citep[][]{Hashimoto2018, Bradac2024}. Similarly to \citet{Bradac2024}, we injected in our low-resolution data the \Lyalpha emission line with an integrated (lensed) flux of 4.3 $\pm$ 1.1 $\times$ 10$^{-18}$ erg s$^{-1}$ cm$^{-2}$, as measured by \citet{Hashimoto2018}. Based on the assumed flux estimate the line should be detected in our prism data at a S/N $\sim$ 9. However, we do not detect any emission line at the expected wavelength (see Fig. ~\ref{fig:grating_prism_map}) and thus conclude that there is no evidence for \Lyalpha emission \citep[][see also]{Bradac2024}. 

Fig. \ref{fig:sfh} shows the SFH derived for each spaxel and for the integrated JD1-N and JD1-S clumps. The SFH highlights that JD1-N is a very recently formed component, with the majority of the stellar mass formed within the last 10 Myr, thus characterised by a very recent peak of SF. Similarly, despite the flatter SFH of JD1-S we observed a strong peak of SF at the source redshift, followed by a flat SFH between z = 9.2 and 10.6, and then a significant decrease to higher redshift. At variance with \cite{Bradac2024}, we do not detect any secondary peak of SF at high redshift for any clump in JD1. Despite this, overall our SFH is qualitatively consistent with previous analysis, showing a recent SF burst \citep{Laporte2021, Bradac2024}.
\begin{table*}
    \begin{center}
    \caption{Summary of the parameters, prior probabilities and posterior probabilities of the fiducial SED fitting \texttt{prospector} model (see also Fig.~\ref{fig:stellar_mass_prospector}). The values reported here are not corrected for lensing magnification. 
    }\label{t.prospector}
    \setlength{\tabcolsep}{4pt}
    \begin{tabular}{llcllc}
  \hline
   & Parameter & Free & Description & Prior & Posterior$_{\rm JD1}$ \\
   & (1)       & (2)  & (3)         & (4)   & (5)  \\
   \hline
   \multirow{11}{*}{\rotatebox[origin=c]{90}{Free parameters}}
   & $\log \mstar [\msun]$ & Y & total stellar mass formed & $\mathcal{U}(7, 13)$ & $8.47^{+0.05}_{-0.05}$ \\
   & $\log Z [\zsun]$ & Y & stellar metallicity & $\mathcal{U}(-2, 0.19)$ & $-1.62^{+0.10}_{-0.10}$  \\
   & $\log \mathrm{SFR}$ ratios & Y & ratio of the $\log \mathrm{SFR}$ between adjacent bins of the SFH & $\mathcal{T}(0, 0.3, 2)$ & ---\\
   & $\tau_V$ & Y & optical depth of the diffuse dust & $\mathcal{G}(0.3,1;0,2)$ & $0.63^{+0.08}_{-0.10}$ \\
   & $\mu$ & Y & ratio between the optical depth of the birth clouds and $\tau_V$ & $\mathcal{U}(-1.0,0.4)$ & $0.02^{+0.05}_{-0.02}$ \\
   & $\sigma_\mathrm{gas} \; [\kms]$ & Y & intrinsic velocity dispersion of the star-forming gas$^\ddag$ & $\mathcal{U}(0,300)$ &  $64^{+44}_{-38}$ \\
   & $\log Z_\mathrm{gas} [\zsun]$ & Y & metallicity of the star-forming gas & $\mathcal{U}(-2, 0.19)$ & $-0.79^{+0.05}_{-0.05}$ \\
   & $\log U$ & Y & ionisation parameter of the star-forming gas & $\mathcal{U}(-4, -1)$ & $-1.02^{+0.01}_{-0.02}$ \\
   \hline
   \multirow{2}{*}{\rotatebox[origin=c]{90}{Other}}
   & $\log SFR_{10} [\msun \, \peryr]$ & N & star-formation rate averaged over the last 10~Myr & --- & $1.46^{+0.04}_{-0.05}$  \\
   & $\log SFR_{100} [\msun \, \peryr]$ & N & star-formation rate averaged over the last 100~Myr & --- & $0.47^{+0.04}_{-0.05}$  \\
  \hline
  \end{tabular}
  \end{center}
(1) Parameter name and units (where applicable). (2) Only parameters marked with `Y' are optimised by \texttt{prospector}; parameters marked with `N' are either tied to other parameters (see Column~4), or are calculated after the fit from the posterior distribution (in this case, Column~4 is empty). (3) Parameter description. For the dust attenuation parameters $n$, $\tau_V$ and $\mu$ see \citet[][their eq.s~4 and~5]{Tacchella2022}. (4) Parameter prior probability distribution; $\mathcal{N}(\mu, \sigma)$ is the normal distribution with mean $\mu$ and dispersion $\sigma$; $\mathcal{U}(a, b)$ is the uniform distribution between $a$ and $b$; $\mathcal{T}(\mu, \sigma, \nu)$ is the Student's $t$ distribution with mean $\mu$, dispersion $\sigma$ and $\nu$ degrees of freedom; $\mathcal{G}(\mu, \sigma, a, b)$ is the normal distribution with mean $\mu$ and dispersion $\sigma$, truncated between $a$ and $b$.
(5) and (6) Median and 16\textsuperscript{th}--84\textsuperscript{th} percentile range of the marginalised posterior distribution for the north and south clump, respectively; for some nuisance parameters we do not present the posterior statistics (e.g., log SFR ratios). $^\ddag$ The velocity dispersion of the emission lines is a nuisance parameter, due to the low spectral resolution of the prism data ($\sim 500$~\kms at the wavelength of \OIIIL); indeed, the posterior probability distributions for JD1 is both consistent with 0~\kms.
\end{table*}
\begin{figure*}
	\includegraphics[width=\linewidth]{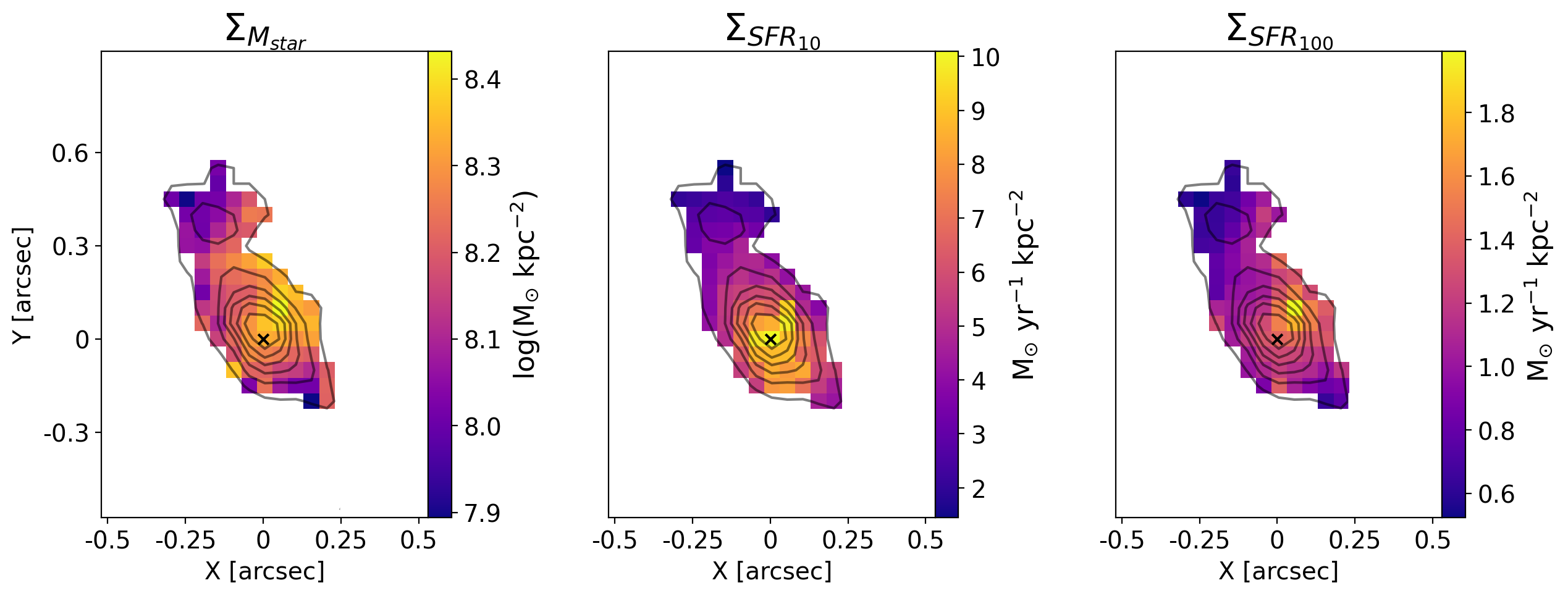}
    \caption{JD1 properties derived from SED fitting of the low-resolution data cube with \textit{prospector} (see Sec. \ref{Subsec3_continuum_fitting}) not corrected for the gravitational lensing effect. From left to right the surface density of the total stellar mass formed, the SFR within the last 10 Myr and 100 Myr. Black contours are arbitrary flux levels from the prism data cube collapsed over the NIRCam F200W filter.}\label{fig:stellar_mass_prospector}
\end{figure*}
\subsection{Gas-phase metallicity}\label{Subsec3_metallicity}
In Sec. \ref{Subsub3_electron_den_temperature} we derived spatially resolved estimates of the ISM electron density and temperature for JD1-S from \OIIL[3726]/\OIIL[3729] and \OIIIL[4363]/\OIIIL line ratios, respectively (see left panels in Fig. \ref{fig:density_temperature_electrons}).
Then, we used the \texttt{getIonAbundance} routine of \textsc{pyneb}, to compute the ionic abundances of the oxygen ions O$^{2+}$ and O$^{+}$ given their respective electron temperatures, gas densities, and the flux ratio of \OIIIL/\Hbeta and (\OIIL[3726]+\OIIL[3729])/\Hbeta. Here, we approximate the oxygen abundance O/H as (O$^{2+}$ + O$^{+}$)/H, neglecting higher ionisation species of O and assume that it is representative of the total gas metallicity \citep[for a detailed discussion on the underlying assumptions of this so called T$_\mathrm{e}$ method see][]{Maiolino2019}. The third panel in Fig. \ref{fig:density_temperature_electrons} shows the direct metallicity map of spaxels with resolved \OII doublet and \OIIIL[4363] emission line measurements. The average value of the gas-phase metallicity of JD1-N and JD1-S is computed from integrated spectra of the selected aperture shown in Fig. \ref{fig:grating_prism_map}. We computed 12 + log(O/H) = 8.3$\pm$ 0.1 and 7.71 $\pm$ 0.25, for the North and South apertures, respectively, consistently on average with the resolved values (see Fig. ~\ref{fig:density_temperature_electrons}).

The right panel of Fig. \ref{fig:density_temperature_electrons} shows the metallicity map as derived with  \textsc{prospector} (i.e. using its grid of photionization modelling) from the nebular fitting of the prism cube. Interestingly, we derive consistent values for the gas metallicity derived via \textsc{prospector} photoionization models fitting of the prism spectra and from direct metallicity measurement using the grating spectra. As a more quantitative comparison, from the prism SED fitting we computed integrated gas metallicities for JD1-N and JD1-S of log(O/H) = 8.00$^{+0.02}_{-0.013}$ and 7.85$^{+0.03}_{-0.03}$, respectively, which are only slightly below (for JD1-N) and above (for JD1-S) the result from the direct method (see Fig. ~\ref{fig:density_temperature_electrons} and Sec. \ref{Subsec3_continuum_fitting}). The metallicity derived via prism fitting of the integrated spectrum of JD1 is log(O/H) = 7.89$^{+0.05}_{-0.05}$, consistent with the value of 7.90$^{+0.04}_{-0.05}$ reported by \citet{Stiavelli2023}. Nevertheless, we observe a North-South gradient for the metallicity content from Fig. \ref{fig:density_temperature_electrons}. JD1-N appears to be more metal enriched compared to JD1-S.

Then, we used \textsc{pyneb} to compute the Neon abundance and derived spatially resolved maps of Log(Ne/O) and Log(Ne/H), as shown in Fig. \ref{fig:neon_abundance}, from \NeIIIL[3869], \Hbeta and \OIIIL emission lines. Our measurements show higher abundances of Ne/H towards JD1-N. Similarly, the Ne/O abundance has a peak co-spatial to the Ne/H, without showing a second maximum towards JD1-N. \citet{Izotov2006} showed a correlation between log(Ne/O) and the gas-phase metallicity, claiming that in high-metallicity H\,\textsc{ii} regions the oxygen is depleted onto dust grains. Since JD1 is supposed to be almost dust free we should not expect a trend of log(Ne/O) with the gas-phase metallicity. Overall, our estimates of the Ne abundances are consistent with previous analysis \citep{Stiavelli2023}.

Many works have shown evidence of a three-dimensional relationship between M$_\star$-Z$_\mathrm{gas}$-SFR that does not appear to evolve with redshift \citep{Tremonti2004, Mannucci2010, Maiolino2019, Cresci2019}. Recently, \citet{Baker2023} have provided evidence of a spatially resolved version of this so-called Fundamental Metallicity Relation (FMR), showing that the local metallicity depends primarily on $\Sigma_{M_{\star}}$ and anti-correlates with $\Sigma_{SFR}$.
We employed the resolved maps of SFR$_{100}$ density and the gas-phase metallicity derived from \textsc{prospector} to test such observed anti-correlation at z$\geq$ 9. Fig.~\ref{fig:anticorr_met_sfr} shows the observed resolved anti-correlation between the gas phase metallicity and the SFR$_{100}$ in JD1. As shown in Fig.~\ref{fig:anticorr_met_sfr}, we performed a Spearman correlation test and found a correlation coefficient $\rho$ = -0.3 and p-value = 0.001, clearly indicating a significant, but mild, correlation between the SFR$_{100}$ density and the gas metallicity. Overall, our findings are qualitatively consistent with the effect of a stream of gas inflowing towards JD1-S from JD1-N (see Sec. \ref{Subsec3_kinematic}). In this scenario, the inflowing gas dilutes the metal content of the ISM, decreasing the metallicity (see right panels in Fig. \ref{fig:density_temperature_electrons}) and replenishing the gas reservoir of JD1-S, 
thus increasing the SFR (see Figs. \ref{fig:stellar_mass_prospector}, \ref{fig:EW_oiii_hb}). Similarly, \cite{Arribas2023} found a prominent north-south metallicity gradient in a z $\sim$ 6.9 galaxy and interpreted it as the result of accretion of metal poor gas from the circumgalactic medium (CGM) into the galaxy \citep[see also][]{Dekel2009, Cresci2010, delpino2024}. Similar results were also found for a sample of compact galaxies studied with NIRSpec/MSA \citep{Tripodi2024}. 

\begin{figure}
	\includegraphics[width=\linewidth]{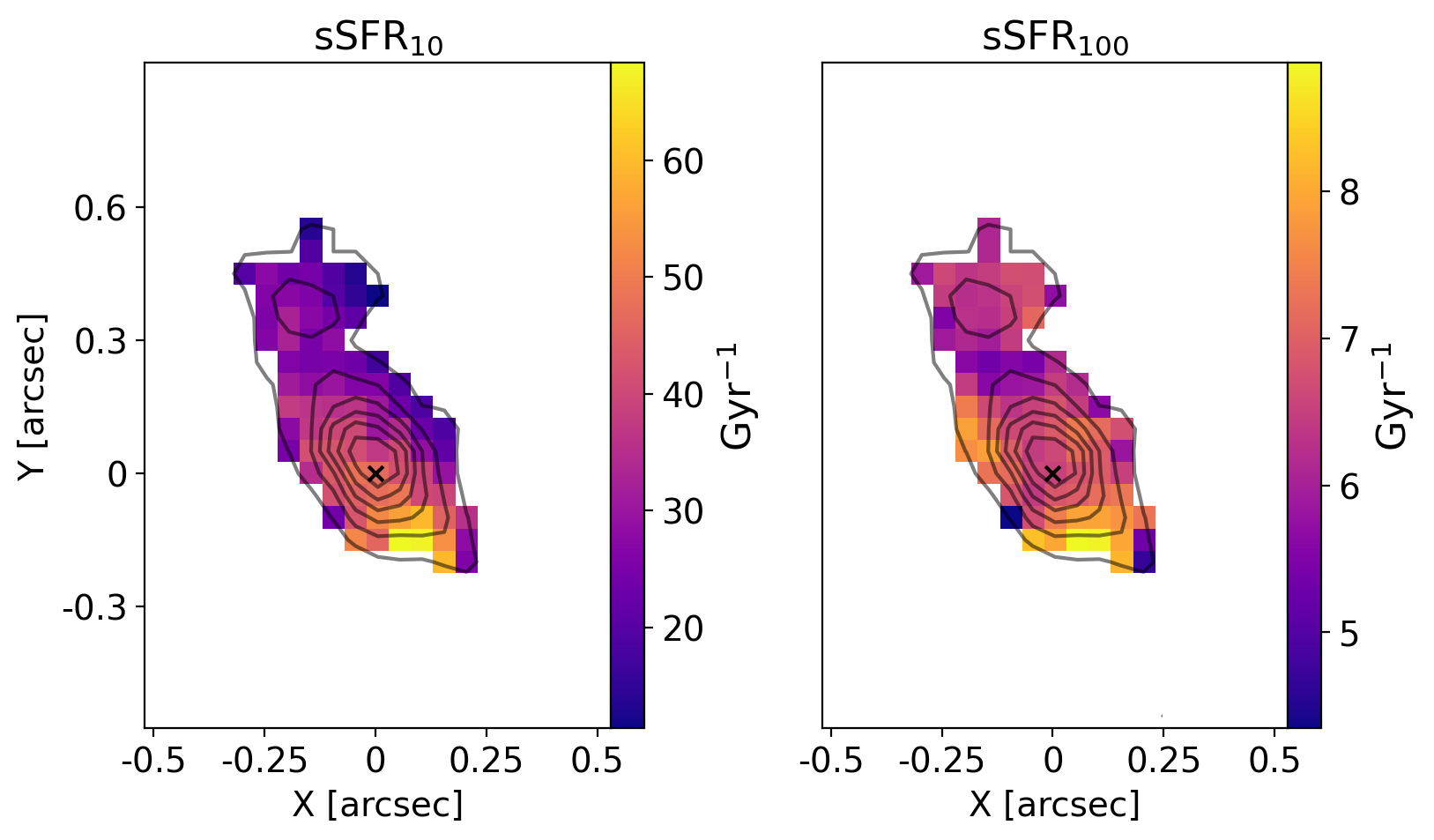}
    \caption{Specific Star formation rate (sSFR) within the last 10 (Left) and 100 (Right) Myr of JD1 not corrected for the gravitational lensing effect. Black contours are arbitrary flux levels from the prism data cube collapsed over the NIRCam F200W filter, as in Fig. ~\ref{fig:stellar_mass_prospector}.}
    \label{fig:specific_SFR}
\end{figure}
\begin{figure}
	\includegraphics[width=\linewidth]{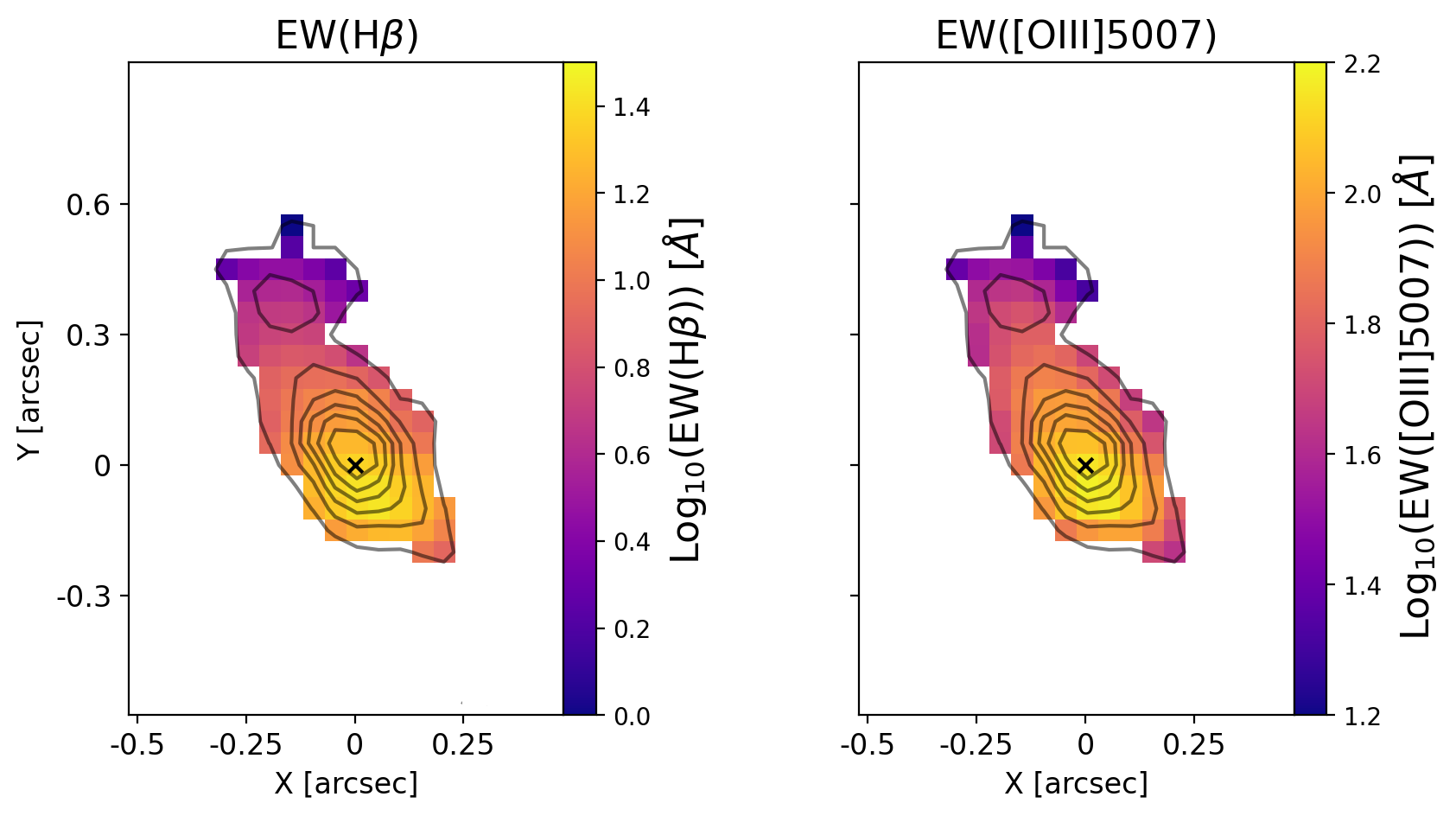}
    \caption{Equivalent width maps of \Hbeta (left) and \OIIIL (right) in logarithmic scale. Black contours are arbitrary flux levels from the prism data cube collapsed over the NIRCam F200W filter, as in Fig. ~\ref{fig:stellar_mass_prospector}}
    \label{fig:EW_oiii_hb}
\end{figure}
\begin{figure}
	\includegraphics[width=\linewidth]{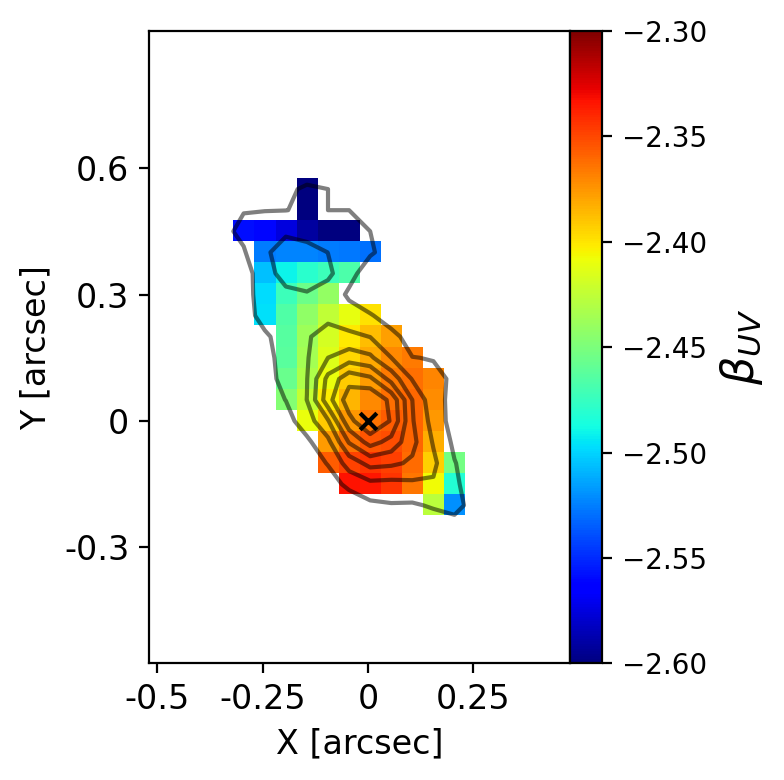}
    \caption{Continuum slope map derived from prism data cube fit in the spectral rest-frame range 1200 to 1500 $\AA$ (see Sec. \ref{Subsec3_continuum_fitting}). Black contours are arbitrary flux levels from the prism data cube collapsed over the NIRCam F200W filter.}
    \label{fig:beta_slope}
\end{figure}

\section{Discussion}\label{Sec4_discussion}
In this paper we presented the spatially resolved properties of JD1, made possible thanks to the combination of high spectral resolution and broad spectral coverage of the G395H and prism NIRSpec/IFS data. Here we summarise and discuss the connection between the north and south companions, JD1-N and JD1-S, also by comparing our results with previous works. 
\subsection{Structure and interaction between the JD1 components}\label{Subsec_north_sout_comps}
Fig. \ref{fig:grating_prism_map} clearly shows two different components in the collapsed continuum, separated by $\sim$ 2 kpc in projection \citep[][]{Hashimoto2018, Stiavelli2023, marquez2023, Bradac2024}. ALMA, HST, and JWST observations support the hypothesis of two different stellar populations in JD1 \citep{Hashimoto2018, Stiavelli2023}. In particular, the most credited scenario is that about half of the current stellar mass of JD1-S formed $\sim$130 Myr before the observed epoch. Then, the galaxy gas reservoir has been replenished by inflows that lead to an ongoing secondary burst of SF, which boosted the observed \OIIIL[88]{\micron} and \Lyalpha emission. Here we speculate that the secondary burst of SF was triggered by the encounter of JD1-S with JD1-N, which is transferring substantial amounts of gas to sustain the recent SF episode in JD1-S. From the kinematic point of view, this idea is consistent with projected gas kinematics in Fig. \ref{fig:OIII007_mom_maps} and \ref{fig:OIII5007_velchan_maps}. Indeed, as clearly shown in the second panel in Fig. ~\ref{fig:OIII5007_velchan_maps}, we observed a collimated filament of ionised gas connecting the two components, with the highest velocity slightly offset from JD1-N and directed towards south. In this scenario we are witnessing the merger between JD1-N and the more massive JD1-S, with gas being dragged towards the gravitational centre. JD1-N has the highest blue-shifted velocity as it is above the plane of the sky with respect to the major galaxy, with the filament representing accreting gas towards South, still dragged by JD1-N, as supported by channel maps in Fig. \ref{fig:OIII5007_velchan_maps}, thus generating the extended blue-shifted tail. 

The inflowing gas moving from North to South replenishes the gas reservoir of the system, possibly sustaining further SF. The panels in Fig. \ref{fig:stellar_mass_prospector} show a peak of the stellar mass density and the SFR density within the last 100 Myr co-spatial with the enhanced velocity dispersion observed in Fig. \ref{fig:OIII007_mom_maps}. This peak is slightly offset from the peak of the \OIIIL emission line and is likely indicating that recent SF is powering outflowing gas, thus increasing the gas velocity dispersion or alternatively is due to beam smearing. 
\begin{figure}
	\includegraphics[width=\linewidth]{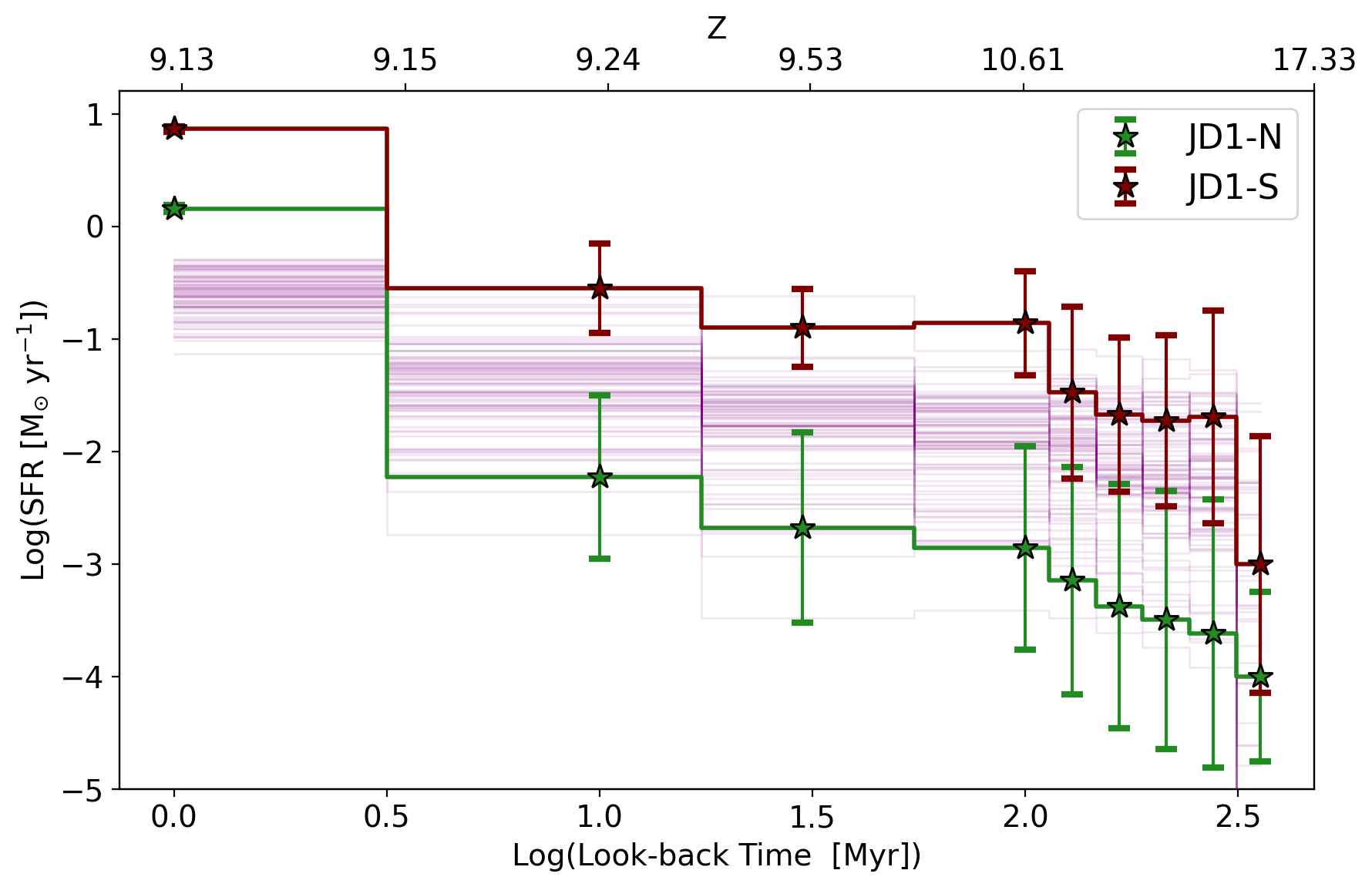}
    \caption{Star formation histories of single spaxels (purple) and integrated JD1-N (green) and JD1-S (maroon) clumps. Errorbars for the main clumps represent the 84\textsuperscript{th} and 16\textsuperscript{th} percentiles.}
    \label{fig:sfh}
\end{figure}
\begin{figure}
	\includegraphics[width=\linewidth]{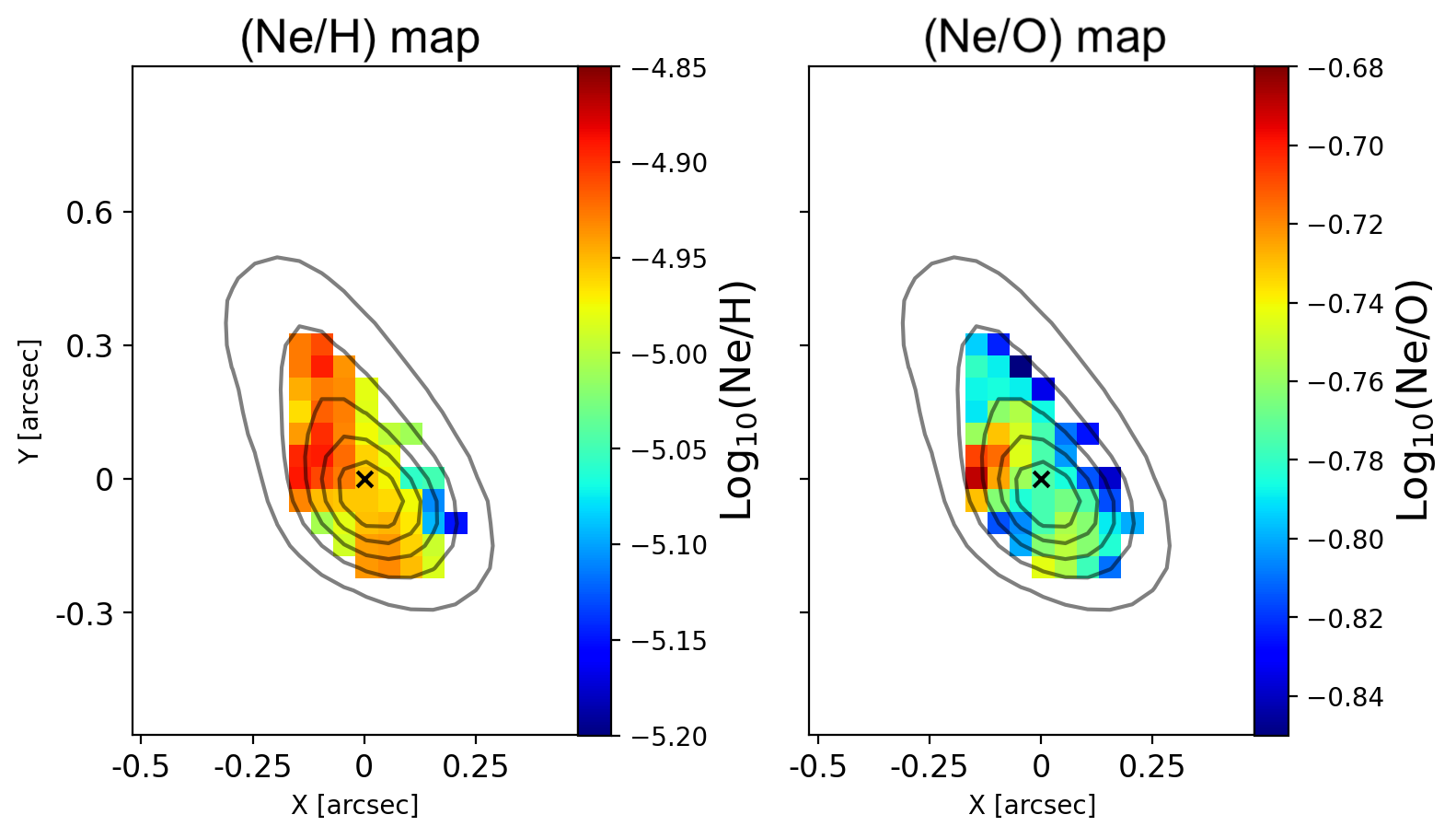}
    \caption{Neon abundance in JD1 with respect to Hydrogen (left) and Oxygen (right). Black solid lines are arbitrary \OIIIL levels. Spaxels with S/N $\le$ 3 are masked.}
    \label{fig:neon_abundance}
\end{figure}

Interestingly, the right panel in Fig. \ref{fig:density_temperature_electrons} shows a peak of the gas-phase metallicity over JD1-N (12 + log(O/H) $\sim$ 7.95), consistent with the integrated value obtained from the direct method. The higher metallicity and \OIIIL/\Hbeta line ratio support the idea of a SF burst over JD1-N which is enriching the ISM with newly forming metals. In particular, many works show that high values of \OIIIL/\Hbeta can be produced by high levels of  sSFR (see Fig. \ref{fig:specific_SFR}), which in turn enhances the ionisation parameter and thus indicates high SF efficiency \citep[][]{Kewley2016, Sanders2016, Dickey2016, Reddy2023}. Finally, the low sSFR observed over the peak of the stellar mass within the last 100 Myr (see Fig. ~\ref{fig:specific_SFR}) indicates that most of the mass of the system must be older than 100 Myr, consistent with previous analysis \citep{Hashimoto2018, Stiavelli2023, marquez2023, Bradac2024}. Our findings are consistent with a younger generation of stars distributed over JD1-N, and an older stellar population in JD1-S raised by the replenishing of gas dragged by this companion (see Fig. \ref{fig:stellar_mass_prospector}, \ref{fig:specific_SFR}). This scenario is also supported by VLT/X-Shooter observations of a diffuse \Lyalpha bubble, supposedly inflated by UV emission of the older population in the JD1-S \citep{Hashimoto2018}.
\begin{figure}
	\includegraphics[width=\linewidth]{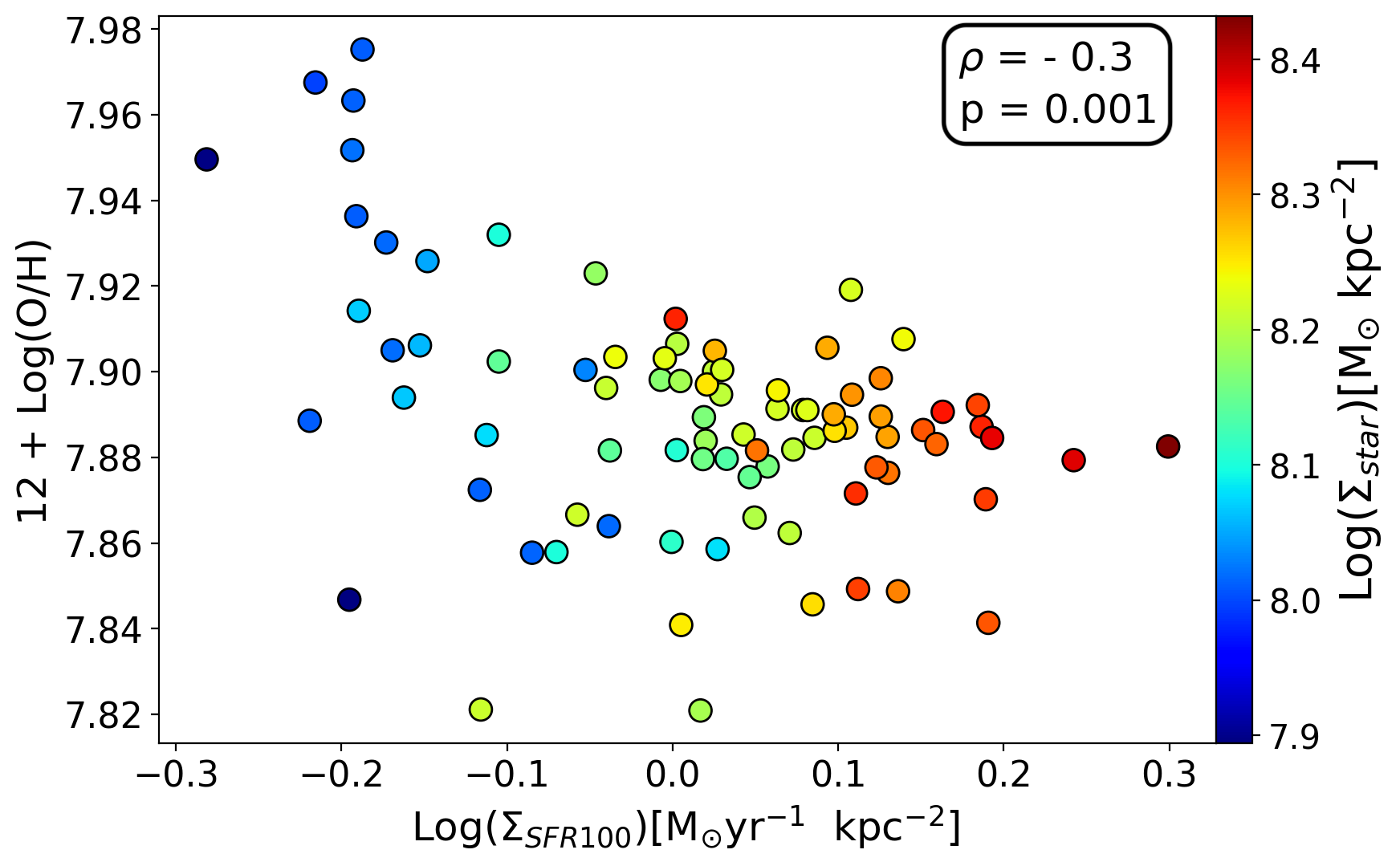}
    \caption{Distribution of the spatially resolved gas-phase metallicity and SFR density colour-coded with the surface density of the total stellar mass formed in JD1. The SFR density and stellar mass surface density are not corrected for magnification effects.  All properties are derived with \textsc{prospector} (see Sec. \ref{Subsec3_continuum_fitting} and Figs. \ref{fig:density_temperature_electrons}, \ref{fig:stellar_mass_prospector}).}
    \label{fig:anticorr_met_sfr}
\end{figure}
\subsection{Resolved electron density at z = 9.11}\label{Subsec_discussion_ISM_proeprties}
\begin{figure*}
	\includegraphics[width=\linewidth]{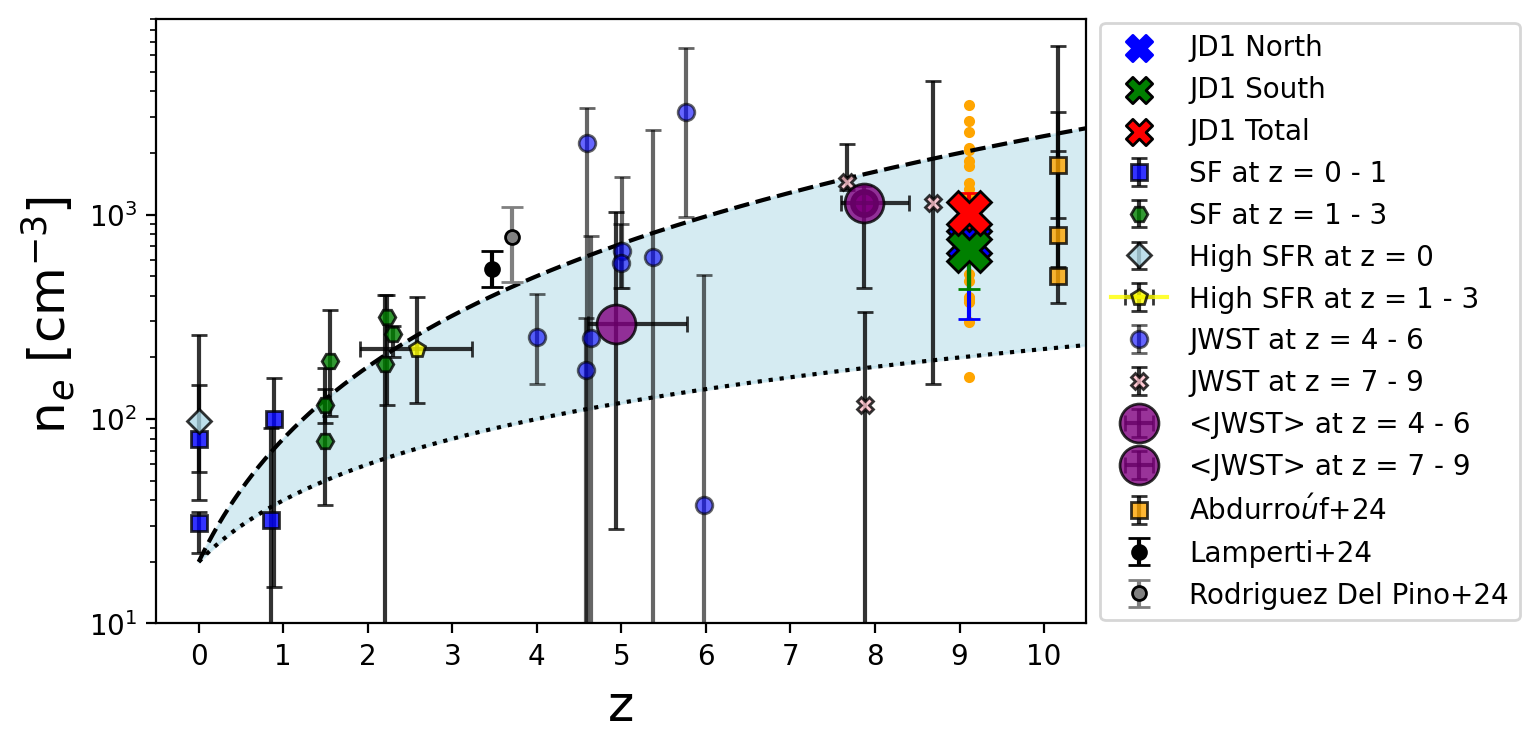}
    \caption{Electron density n$_\mathrm{e}$ as a function of redshift, adapted from \citet{Isobe2023} (see references therein). Dotted and dashed curves represent n$_\mathrm{e}$ following the $\propto$(1 + z) and $\propto$(1 + z)$^{2}$ relations, respectively. Blue, green and red crosses represent the integrated values for JD1-N, JD1-S and for the total integrated emission, respectively. Small orange circles represent the electron density of single spaxels as derived from our spatially resolved analysis. Blue squares and green hexagons show median n$_\mathrm{e}$ of SFMS galaxies at z $\sim$ 0–1 \citep{Berg2012, Swinbank2019, Davies2021} and z $\sim$ 1–3 \citep{Steidel2014, Sanders2016, Kaasinen2017, Kashino2017, Davies2021}, respectively. Cyan diamond and yellow pentagon show the average n$_\mathrm{e}$ from high specific SFR galaxies at z =0 \citep{Berg2022} and z =1-3 \citep{Christensen2012a, Christensen2012b, Sanders2016b, Gburek2019}, respectively. Small blue circles and pink crosses show single \JWST galaxies at z $\sim$ 4–6 and z $\sim$ 7–9, respectively \citep{Isobe2023}. The large purple circle and double circle denote the 50th and 16th–84th percentiles of the properties of the \JWST galaxies at z $\sim$ 4–6 and z $\sim$ 7–9, respectively. Orange squares show n$_\mathrm{e}$ derived from single clumps and stacked spectra of a single \JWST z$\sim$ 10.1 lensed galaxy from \citet{Abdurrouf2024}. The black and grey circles are electron density estimates at z $\sim$ 3.5 by GA-NIFS observations \citep{delpino2024, Lamperti2024}.  }
    \label{fig:density_with_z}
\end{figure*}

Taking advantage of the high spatial and spectral resolution of the G395H data cube, we estimated the electron density from the integrated spectra of the two main companions in JD1 and provide the first spatially resolved electron density map at high redshift of the main galaxy. Fig. \ref{fig:density_with_z} shows a collection of electron density estimates from the local Universe up to redshift 8.68 compiled by \citet{Isobe2023} with the addition of electron density estimates at z $\sim$ 3.5 by the GA-NIFS project \citep{Lamperti2024, delpino2024} and z$\sim$ 10.1 by \citet{Abdurrouf2024}. The black dashed and dotted lines represent the n$_\mathrm{e}$ $\propto$ $\times$ (1+z)$^{p}$ relation, with p = 2 and 1, respectively. Our estimates for the electron density over the integrated JD1-N, JD1-S components, as well as the total integrated value for JD1 and the values derived from the spatially resolved analysis, represented as scattered blue points, are in agreement with the predicted trend. In particular, by using high-z JWST observations and extrapolating the trend of low-z galaxies up to z $\sim$ 9, \citet{Isobe2023} suggest that the electron density up to z = 9 is well fitted by n$_\mathrm{e}$ $\propto$ $\times$ (1+z)$^{p}$, with p = 1 - 2. A possible scenario to explain the observed trend \citep{Isobe2023, Abdurrouf2024} relates both to the decrease of diffused metals \citep{Davie2011} into the ISM and to the redshift evolution of galaxy sizes with redshift \citep{Kartaltepe2023, Ormerod2024}. In particular, as suggested by \citet[][see also \citealp{Davies2021}]{Isobe2023}, the gas density of SF regions at high redshift are higher with respect to local analogues due to their higher compactness, due to the galaxy sizes scaling with redshift as R $\propto$ (1+z)$^{-n}$, with $n \sim$ 0.7 - 1 \citep{Shibuya2015, Ono2023, Ormerod2024}. Assuming a disc-like configuration with redshift-independent thickness, we would get n$_\mathrm{e}$ $\propto$ R$^{-2}$ which gives the n$_\mathrm{e}$ $\propto$ $\times$ (1+z)$^{1.4\text{-}2}$ relation. On the other hand, recent works motivate the n$_\mathrm{e}$ $\propto$ $\times$ (1+z)$^{1}$ relation as mainly due to the metallicity evolution with redshift \citep{Abdurrouf2024}. In particular, hydrodynamic simulations of high-z galaxies revealed a clear anti-correlation between the mean cloud density and the gas metallicity ($Z$) described by $\Bar{n} \  \propto Z^{-1}$ \citep{Garcia2023}, as due to the high temperatures and low metallicity of high redshift systems \citep{Kewley2019}. Combining the mean gas density evolution with the observed and predicted gas metallicity trend with redshift: $Z \propto(1+z)^{-1}$ \citep{Davie2011, Yuan2013}, we get the dotted relation: n$_\mathrm{e}$ $\propto$ $\times$ (1+z)$^{1}$.

\subsection{On the rotating disc nature and the third clump of JD1}\label{Subsec_discussion_rotating_nature}
\begin{figure}
	\includegraphics[width=\linewidth]{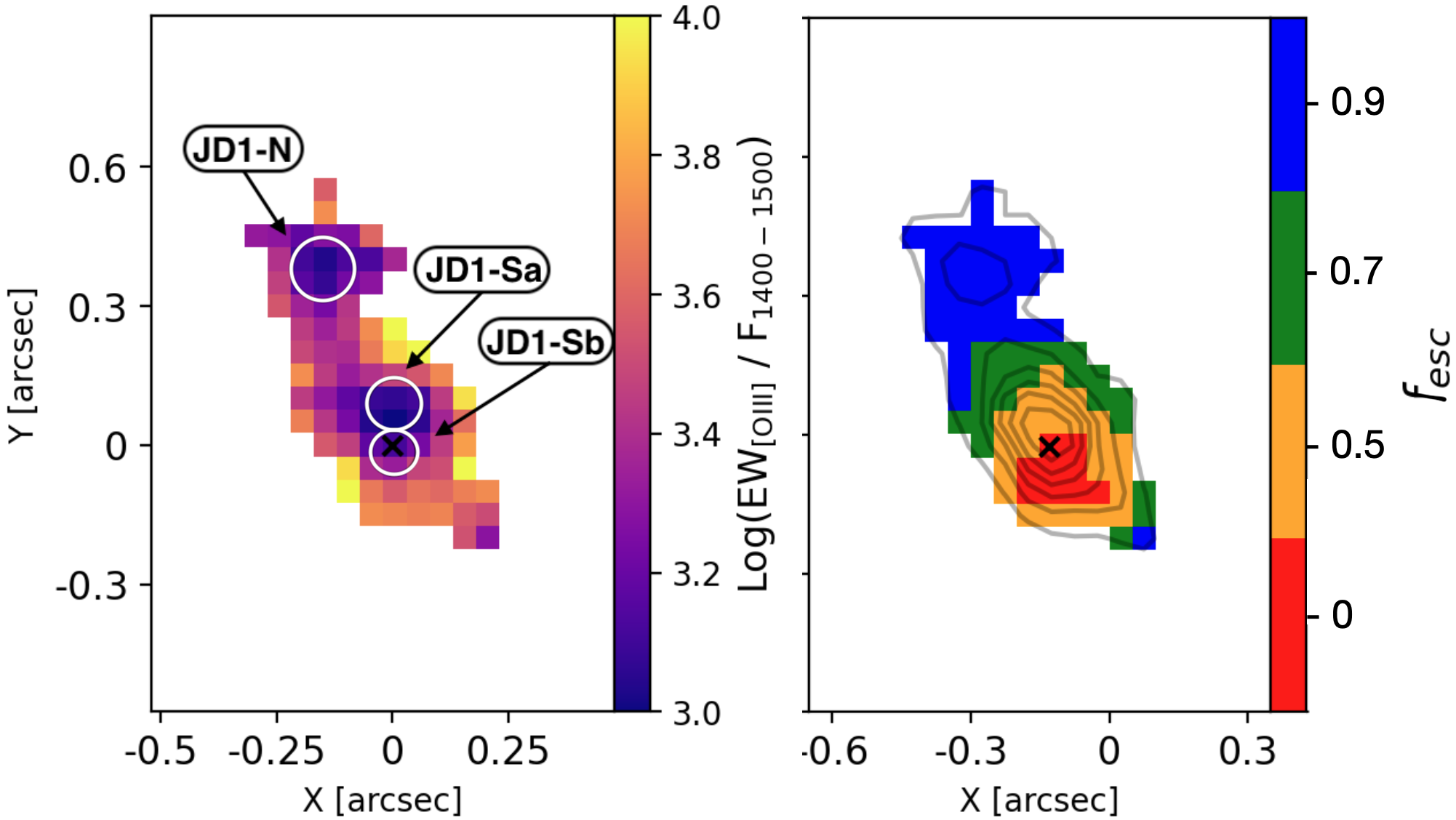}
    \caption{Left: Ratio between the \OIIIL EW and the continuum between 1400 \micron and 1500 \micron rest frame. The black cross marks the peak of the \OIIIL emission line. The three photometric unresolved clumps are highlighted by white circles. Right: Escape fraction map derived from the relations in \citet{Zackrisson2017}. Black contours are arbitrary flux levels from the prism data cube collapsed over the NIRCam F200W filter.}
    \label{fig:O3_minus_cont14001500}
\end{figure}

ALMA \OIIIL[88\micron] observations of JD1 show a smooth gas distribution, with no resolved clumps within the system. \cite{Tokuoka2022} computed spatially resolved kinematics of the \OIII[88\micron] line that shows a smooth north--south velocity gradient, as we observed from \OIIIL emission line, and concluded that  the best scenario for JD1 is a rotating disc galaxy.
JWST NIRCam F150W and F200W observations bring evidence of two smaller clumps building up the main JD1-S clump \citetext{JD1-Sa and JD1-Sb, see Fig. ~\ref{fig:O3_minus_cont14001500}; \citealp{Stiavelli2023, Bradac2024}}. In particular, \citet{Bradac2024} identified three unresolved star-forming clumps with an underlying smooth galaxy component.
From our analysis, the high-resolution data does not allow us to resolve the JD1-N clump, which instead is well detected in the low-resolution data. Interestingly, we observe that the position of JD1-Sb coincides with the peak of the \OIIIL emission line, whereas JD1-Sa is shifted towards North by $\approx$ 450 pc, thus approximately located along the enhanced velocity dispersion region identified by \OIIIL emission line (see Fig. \ref{fig:OIII007_mom_maps}). We investigated the role of JD1-Sa and JD1-Sb in the big picture of the kinematic in JD1 to test whether the disc nature of the system can be confirmed or not. Assuming that JD1-Sb is the centre of the rotation is actually inconsistent with the observed kinematics, as there is no positional agreement with the peak of velocity dispersion shown from \OIIIL analysis (see Fig. \ref{fig:OIII007_mom_maps}).
Additional insights come from the analysis of the escape fraction of ionizing photons.
Recent works show evidence of high escape fraction of UV photons in bright \OIIIL line emitters \citep{Izotov2018, Fletcher2019, Nakajima2020}. Therefore, we compared the equivalent width of the \OIIIL emission line with the UV continuum to estimate the fraction of ionising photons produced by young stars that can escape the galaxy, without being absorbed by the diffuse ISM. We expect higher escape fractions in regions with lower emission line to continuum ratio and steeper continuum. The left panel in Fig. \ref{fig:O3_minus_cont14001500} shows the ratio of the \OIIIL EW and UV continuum emission between 1400 and 1500 \AA \ rest frame. Overall, we observe deep minima of the ratio to be coincident with the location of JD1-N and JD1-Sa, and steeper UV slope ($\beta_{UV}$) with respect to JD1-Sb (see Fig. \ref{fig:beta_slope}). The right panel in Fig. \ref{fig:O3_minus_cont14001500} shows the escape fraction map of JD1, derived interpolating the predicted distribution of the escape fraction as a function of EW(H$\beta$) and the UV slope $\beta_{UV}$ for simulated dust-free galaxies between z = 7 - 9 \citep{Zackrisson2017}. While the precise value of these estimates is model dependent, they show a clear trend of increasing f$_{\rm esc}$ away from JD1-Sb, including towards JD1-Sa and JD1-N. Under the hypothesis of a highly inclined disc galaxy, as suggested by the global morphology, it is unlikely that the gravitational centre is associated with higher levels of escape fraction with respect to the galaxy edges. 
Finally, the absence of dust \citep[][]{Hashimoto2018} brings additional evidence of a merger scenario instead of the rotating disc scenario; this is because an edge-on disc -- particularly with the high gas fraction we inferred from the dynamical mass estimate -- may have relatively high dust column densities, even at low metallicity. Instead, there is no evidence for dust from the region with high velocity dispersion, which would represent the centre of the disc and, therefore, the highest column densities.
\section{Conclusions}\label{Sec_conclusions}

In this work, we used integral-field spectroscopy data from \JWST NIRSpec/IFS to perform a detailed spatially resolved analysis of MACS1149-JD1, a bright, lensed galaxy at z=9.11.
We combined the high spectral resolution of the G395H grating with the excellent sensitivity and broad spectral coverage of the prism to draw a comprehensive picture of this system.
We studied the ionised gas kinematics and excitation properties through emission-line fitting, and we performed a spaxel-by-spaxel full-spectrum fitting analysis of the prism and obtained the stellar mass and gas metallicity. The main results are summarised as follows:
\begin{itemize}
    \item The emission line multi-Gaussian fit of the high-resolution G395H data cube revealed the peculiar ionised gas kinematics of this system. Velocity channel maps in Fig. \ref{fig:OIII5007_velchan_maps} show a smooth velocity gradient between JD1-S and JD1-N, with a peak of projected velocity of $\sim$ -200 km s$^{-1}$ in correspondence of JD1-N. The spatially resolved gas velocity dispersion map shows a maximum of 90 km s$^{-1}$ perpendicular to the direction of the blue tail, $\sim$ 0.8 kpc NW from JD1-S. 

    \item We computed the electron density and temperature from spatially resolved emission line ratios of \OIIall doublet and \OIIIL/\OIIIL[4363] (see Fig. \ref{fig:density_temperature_electrons}). From integrated spectra of JD1-N and JD1-S we found electron densities of 730 cm$^{-3}$ and 670 cm$^{-3}$, respectively (see Fig. \ref{fig:density_with_z}). From direct T$_\mathrm{e}$ measurements, we estimated the oxygen ion abundance and computed the gas-phase metallicity distribution. From integrated spectra of the two clumps we measured 12 + log(O/H) = 8.3 and 7.71, in JD1-N and JD1-S, respectively.

    \item We use recently proposed emission line ratio diagrams tailored to investigate the ionisation source in high-z systems employing \OIIIL/\OIIIL[4363] vs \OIIIL[4363]/\Hgamma, \OIIIL/\OIIL[3727] vs \OIIIL[4363]/\Hgamma, and \NeIIIL/\OIIL[3727] vs \OIIIL[4363]/\Hgamma emission line ratios. Both these diagrams and the standard BPT diagram show no clear AGN signatures, thus everything is well consistent with ionisation from SF.
    
    \item We fit the SED from the low-resolution data cube and provide spatially resolved estimates of the stellar mass budget, SFR and sSFR at different epochs, and gas-phase metallicity. We confirm the presence of two distinct populations, with JD1-N being characterised by a recent SF burst and the major contribution to the stellar mass to be ascribed to an older population distributed over JD1-S (see Fig. \ref{fig:sfh}), co-spatial with the enhancement of the ionised gas velocity dispersion (Fig. \ref{fig:OIII007_mom_maps}).

    \item From SED fitting and direct-method analysis we measured a north-south gradient of gas-phase metallicity, with higher values over JD1-N. We provided first evidence of a resolved anti-correlation between the metallicity and SFR density at such high redshift. We motivate the observed inverse correlations as due to an inflow of gas over JD1-S from JD1-N, which is diluting the ISM metal content and boosting SF.

    \item As for the nature of JD1, we weigh both the merger and rotating disc hypotheses. The gas kinematics are broadly consistent with a disc, albeit with a tilted kinematic axis, and a peak of velocity dispersion offset from the morphological centre. In this case, JD1 would have
    M$_{\rm dyn}$ = 1.2$^{+0.5}_{-0.4}$ $\sqrt{10/\mu}$ $\times$ 10$^{9}$\msun
    and a stellar-to-total mass fraction of 8 percent. However, this scenario does not explain the smooth metallicity gradient (Fig.~\ref{fig:density_temperature_electrons}), the negligible dust attenuation, the low gas content, and the high escape fraction near the putative centre of the disc (Fig.~\ref{fig:O3_minus_cont14001500}). All of these observables can instead be naturally explained if JD1-S and JD1-N were two distinct systems interacting.
\end{itemize}

Our findings represent a step forward in the comprehension of the interplay between star formation processes, chemical enrichment and merger at high-redshift.
Moreover, this work highlights the major impact of \JWST/NIRSpec IFS observation in constraining the ISM properties and galaxy evolution processes in the early Universe.

\section*{Acknowledgements}

CM, RM and FDE acknowledge support by the Science and Technology Facilities Council (STFC), by the ERC Advanced Grant 695671 ``QUENCH'', and by the UKRI Frontier Research grant RISEandFALL. CM also acknowledge the support of the INAF Large Grant 2022 ``The metal circle: a new sharp view of the baryon cycle up to Cosmic Dawn with the latest generation IFU facilities'' and of the grant PRIN-MUR 2020ACSP5K\_002 financed by European Union – Next Generation EU.
RM is further supported by a research professorship from the Royal Society.
SA, BRdP, and MP acknowledge support from the research project PID2021-127718NB-I00 of the Spanish Mi    nistry of Science and Innovation/State Agency of Research (MICIN/AEI).
H{\"U} gratefully acknowledges support by the Isaac Newton Trust and by the Kavli Foundation through a Newton-Kavli Junior Fellowship.
GCJ, AJB acknowledges funding from the "FirstGalaxies" Advanced Grant from the European Research Council (ERC) under the European Union’s Horizon 2020 research and innovation programme (Grant agreement No. 789056).
SC, EP, and GV acknowledge support by European Union's HE ERC Starting Grant No. 101040227 - WINGS.
GC acknowledges the support of the INAF Large Grant 2022 ``The metal circle: a new sharp view of the baryon cycle up to Cosmic Dawn with the latest generation IFU facilities''

\section*{Data Availability}
The JWST/NIRSpec data used in this work has been obtained within
the NIRSpec-IFU GTO programme (program ID 1262, Observation 7) and are publicly available since June 5, 2024. The data presented in this work will be shared upon reasonable request to the corresponding author.



\bibliographystyle{mnras}
\bibliography{mnras_template.bbl}






\bsp	
\label{lastpage}
\end{document}